\documentclass[sigconf, screen, nonacm]{acmart}

\usepackage{xcolor}
\usepackage{tikz}
\usepackage{pgfplots}
\usetikzlibrary{arrows.meta, positioning, calc}
\usepackage{caption, subcaption}
\usepackage{graphicx}
\usepackage{multicol}
\usepackage{booktabs}  
\usepackage{array}     
\usepackage{enumitem}
\usepackage{xspace}

\usepackage[utf8]{inputenc} 
\usepackage{pgfplots}
\usepackage{pgfplotstable}
\usepackage{adjustbox}
\pgfplotsset{compat=1.18}
\usepackage{fixltx2e}
\usetikzlibrary{shapes.geometric, arrows.meta, positioning}
\usetikzlibrary{shapes, arrows}
\usetikzlibrary{positioning, calc} 
\usetikzlibrary{decorations.pathmorphing} %
\usepackage{multirow}
\usepackage{subcaption}
\usepackage{tcolorbox}

\usepackage{soul}
\usepackage{enumitem}

\newcommand*\wcircled[1]{\tikz[baseline=(char.base)]{
            \node[shape=circle,draw,inner sep=0.4pt,fill=white,text=black] (char) {\texttt{\textbf{#1}}};}}

\newcommand*\circled[1]{\tikz[baseline=(char.base)]{
            \node[shape=circle,draw,inner sep=0.4pt,fill=black,text=white] (char) {\texttt{\textbf{#1}}};}}

\newcommand{\hlyellow}[1]{{#1}}
\newcommand{\hlgreen}[1]{#1}

\newcommand{\hlred}[1]{{#1}}
\newcommand{\hlorange}[1]{{#1}}
\newcommand{\hlpurple}[1]{#1}

\newcommand{\hlbrown}[1]{#1}
\newcommand{\hlgray}[1]{#1}

\newcommand{\hlindigo}[1]{{#1}}

\newcommand{\papertitle}{PRACtical}

\newcommand{\allchanges}[1]{{#1}} 

\definecolor{alertred}{RGB}{200, 0, 0}         
\definecolor{alertblue}{RGB}{0, 102, 204}      
\definecolor{alertgreen}{RGB}{0, 153, 0}       
\definecolor{alertorange}{RGB}{255, 140, 0}    
\definecolor{alertpurple}{RGB}{153, 51, 255}   
\definecolor{alertgray}{RGB}{102, 102, 102}    

\begin{document}

\twocolumn

\title{\papertitle{}:
Subarray-Level Counter Update and Bank-Level Recovery Isolation for Efficient PRAC Rowhammer Mitigation}

\author{Ravan Nazaraliyev\quad Saber Ganjisaffar\quad Nurlan Nazaraliyev\quad Nael Abu-Ghazaleh}
\affiliation{%
  \institution{University of California, Riverside}
  \country{}
}
\email{{rnaza005, sganj003, nnaza008, naelag}@ucr.edu}

\begin{abstract}
 As DRAM density continues to scale, the Rowhammer vulnerability increases in severity due to heightened charge leakage, which lowers the activation threshold required to induce bit flips. To mitigate this risk, industry-standard solutions have shifted from memory controller-based row activation counters, which require large SRAM storage with significant area and power overheads, to in-DRAM row activation counters. The DDR5 JEDEC standard incorporates a modified DRAM architecture featuring per-row activation counters (PRAC) and an Alert Back-Off (ABO) signal that notifies the memory controller (MC) to trigger mitigation mechanisms. However, PRAC introduces a performance overheads by incrementing counters during the precharge operation, adding an additional delay to the precharge phase. Furthermore, when the ABO signal is triggered upon a row reaching the Alert threshold, \texttt{RFM\textsubscript{ab}} indiscriminately stalls all memory requests at the memory channel level, even when only a single bank is being accessed heavily, leading to unnecessary performance degradation.  In this work, we propose \papertitle, an optimized approach to improving the performance of existing PRAC+ABO mechanisms while maintaining security guarantees. To reduce counter update latency, we introduce a centralized increment circuit, allowing the memory controller to proceed with subsequent activations to other subarrays without suffering the increment delays. To minimize unnecessary memory stalls and make the system resilient against memory performance attacks based on channel stalling upon Alert, we enhance \texttt{RFM\textsubscript{ab}} with bank-level granularity, enabling the memory controller to selectively stall only the affected banks rather than the entire memory channel. This is achieved by introducing a register in the DRAM that shows the banks under attack upon an Alert signal.  Our proposed techniques improve the performance over state of the art opportunistic PRAC and ABO, by 8\% (20\% max). Additionally, \papertitle{} saves an average of 19\% energy relative to PRAC+ABO. \papertitle{} is resilient to performance attacks, showing less than 6\% slowdown on an aggressive performance attack, while providing the same security as PRAC+ABO.
\end{abstract}


\maketitle

\sloppy
\section{Introduction}
Rowhammer~\cite{cojocar2019exploiting, de2021smash, mutlu2019rowhammer, kim2014flipping} is a well-known serious vulnerability affecting DRAM: repeated access to one or more target rows can induce bit flips in adjacent rows. The bit flips occur due to charge leakage from repeated activations of a DRAM row, which can accelerate charge leakage in nearby memory cells, causing their state to change. When a row is activated a sufficient number of times within a single refresh interval, it may corrupt the contents of nearby rows. The estimated minimum number of such activations that could induce a bit flip is referred to as the \textit{Rowhammer Threshold}, denoted by $N_{\mathrm{RH}}$. \hlindigo{As DRAM technology has scaled to smaller feature sizes, $N_{\mathrm{RH}}$ has significantly decreased—from approximately 140K activations in earlier DDR3 devices to as low as 4.8K in LPDDR4 modules}~\cite{kim2014flipping, kim2020revisiting}; it is expected to drop further in future generations. Concerningly, the significant drop in $N_{\mathrm{RH}}$ leads to a noticeable increase in the frequency of DRAM bit flips~\cite{loughlin2021stop}. \allchanges{Accordingly, mitigation strategies must continuously evolve to remain effective against increasingly demanding and sophisticated threat conditions.}

DRAM plays a critical role in determining the performance of memory-intensive workloads, leading to the well known memory wall~\cite{wulf-95,gholami-24,ghose2019demystifying}.  As Rowhammer threats continue to increase with memory scaling, the proposed mitigations have come at a substantial cost to DRAM performance.  JEDEC~\cite{jedec2024pressrelease} has recently introduced a new standard mitigation mechanism for the Rowhammer vulnerability in DDR5 memory devices. This mechanism incorporates two key components: \emph{Per-Row Activation Counters} (PRAC)~\cite{jedec2024ddr5} and \emph{Alert Back-Off} (ABO) protocol. PRAC, inspired by the Panopticon framework~\cite{bennett2021panopticon}, implements an in-DRAM architectural enhancement where each row is extended to store an activation counter. \hlindigo{When the number of activations to a row exceeds a predefined Alert threshold, the mechanism should trigger mitigation for that row and its neighboring rows.} This mechanism enables localized, row-level tracking without incurring the area and power overhead associated with external SRAM-based counters~\cite{qureshiautorfm, saxena2024start}. 
The ABO protocol~\cite{jedec2024ddr5} complements PRAC by providing a signaling mechanism through which the DRAM device can notify the memory controller (MC) when a critical activation threshold is crossed. \hlindigo{The MC sends all-bank RFM} (\texttt{\hlindigo{RFM}\textsubscript{\hlindigo{ab}}}) \hlindigo{triggering targeted refresh operations, thereby preventing potential bit flips.} While the PRAC+ABO framework marks a significant improvement in integrated, hardware-supported Rowhammer mitigations under reduced thresholds~\cite{canpolat2025chronus,qureshi2024moat,woo2025qprac}, it also introduces several critical limitations that compromise both performance and scalability, particularly in high-performance or multi-threaded environments.

\begin{figure*}[t]
  \centering
  \includegraphics[width=1\linewidth]{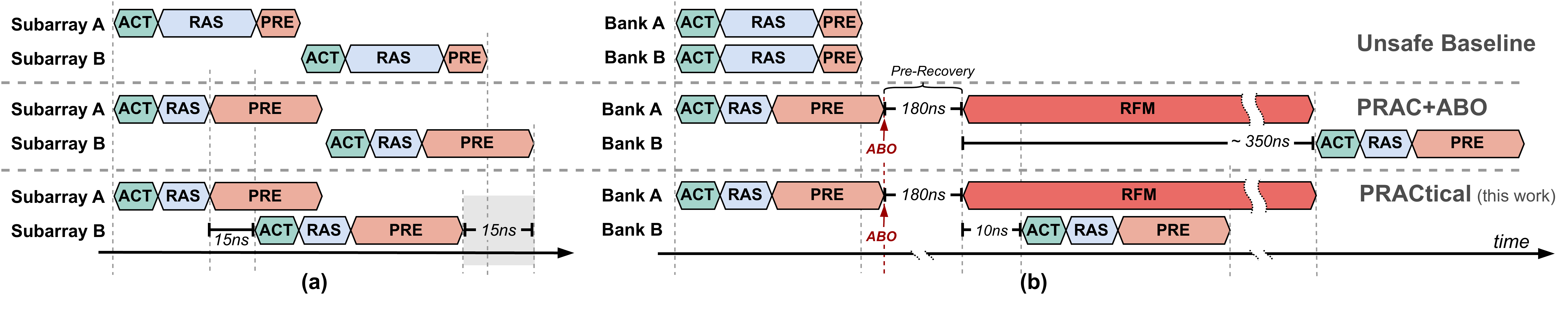}
  \caption{ \allchanges{Comparison of baseline DRAM (top), PRAC+ABO (middle), and the proposed design (bottom). PRAC+ABO introduces (a) access latency from counter update overhead and (b) all-bank stalls due to coarse-grained \allchanges{\texttt{RFM} recovery}. The proposed mechanism mitigates both by enabling subarray-level PRAC updates and bank-level \allchanges{mitigation}. The gray regions illustrate the performance improvements enabled by the proposed approach. Register read operation takes 10ns in \papertitle{} in part (b).} 
  }
  \Description{}
  \label{fig:motiv}
\end{figure*}

First, the integration of PRAC requires architectural modifications that impact critical DRAM timing parameters~\cite{hassan2024self, jedec2024pressrelease}. Specifically, the logic responsible for updating the per-row counters introduces an additional latency of around 5\emph{ns} to the row cycle time ($t_{\mathrm{RC}}$). 
 Even more significantly, the precharge timing ($t_{\mathrm{RP}}$) is extended from \allchanges{15\emph{ns}} to 36\emph{ns} to accommodate the read-modify-write (RMW) operations needed for per-row counter updates. These modifications directly affect memory access latency and throughput, especially in workloads with high row buffer conflicts, as illustrated in Figure~\ref{fig:motiv} (a) (Baseline vs. PRAC+ABO).

Second, \allchanges{after ABO is signalled to MC from DRAM,  MC sends \texttt{RFM\textsubscript{ab}} command to DRAM for recovery. This command operates at the memory channel granularity. Therefore, the memory controller must conservatively stall all requests across the entire channel, even if only a single bank is affected, as illustrated in the middle row of Figure~\ref{fig:motiv} (b)}. \allchanges{Since all banks undergo recovery, even in the absence of an actively hammered (hot) row, the mechanism \textit{opportunistically} refreshes potential victim rows to preemptively mitigate future attacks. Our analysis reveals that this opportunistic strategy results in a threefold increase in recovery refreshes, underscoring the inefficiency and potential redundancy of these additional operations.}
 We further observe that, across a set of memory-intensive benchmarks, on average, only 1.16 out of 64 banks \allchanges{need} recovery at any given time (See \S\ref{sec3.2}). Consequently, the remaining banks are unnecessarily stalled, leading to avoidable performance degradation. This coarse-grained design limits memory-level parallelism, degrading performance and responsiveness in scenarios where finer-grained mitigation would suffice.

While the PRAC+ABO mechanism represents meaningful progress toward an in-DRAM Rowhammer mitigation, it introduces two key performance-related drawbacks: (1) increased memory access latency due to counter update overhead, and (2) channel-wide stalls caused by the coarse granularity of \allchanges{\texttt{RFM\textsubscript{ab}}}. These challenges do not undermine the mechanism’s ability to prevent Rowhammer bit flips, but they do limit its efficiency and suitability for performance-sensitive systems. As such, there is a pressing need for a more fine-grained, low-overhead mitigation framework that minimizes latency and preserves DRAM throughput while maintaining effective protection.
To address the performance and energy limitations of the existing PRAC+ABO framework, we propose \textit{\textbf{\papertitle{}}} — a minimal and efficient redesign that improves both responsiveness and scalability of PRAC+ABO. \hlred{ This new design introduces two key enhancements. First, it leverages subarray-level increment logic to decouple counter updates from global precharge timing, allowing subsequent activations to proceed in other subarrays without incurring the additional latency associated with PRAC’s read-modify-write operations.} \hlred{
While prior works have explored and leveraged subarray-level parallelism to enhance memory performance and efficiency}~\cite{kim2012case,hassan2024self}\hlred{, the introduction of the PRAC standard imposes new constraints that limit such parallelism. In this work, we demonstrate that with modest modifications to the DRAM circuitry, it is possible to restore a degree of subarray parallelism by overlapping the activation of a new row with the counter update of the previously accessed row.}
 \hlred{This significantly reduces access delay and enables} \hlred{known} \hlred{
subarray-level parallelism}  \hlred{
for per-row counter updates}\hlred{. Second,} \hlred{
\papertitle{} provides a new \texttt{RFM} command called \texttt{RFM\_MASK} that eliminates stalling every bank once ABO is signalled from DRAM to MC}
\hlred{, thereby ensuring that only the affected banks are stalled in response to a threshold violation. This fine-grained control minimizes unnecessary interference with unrelated memory traffic and improves overall system throughput. \papertitle{} makes small modifications to the memory controller and DRAM.}

Together, these enhancements enable precise, low-latency and \allchanges{low-energy} Rowhammer mitigation with minimal disruption to normal DRAM operations, as demonstrated in the bottom row of Figure~\ref{fig:motiv} (a) and (b). Our performance evaluation, detailed in Section~\ref{sec:eval}, utilizing DRAM simulation with Ramulator~\cite{luo2023ramulator}, demonstrates that \papertitle{} achieves \allchanges{mean performance improvement of 8\%} over \allchanges{opportunistic} PRAC+ABO for high-performance and memory-intensive applications while maintaining the security guarantees of PRAC+ABO-based mitigations~\cite{canpolat2025chronus,woo2025qprac,qureshi2024moat}.
In summary, the key contributions of this paper are: 
\begin{itemize}[noitemsep, topsep=0pt, leftmargin=1.5em]

\item We identify two major performance bottlenecks in the PRAC+ABO mechanism: precharge-induced latency caused by PRAC updates, and coarse-grained channel-wide stalling due to \allchanges{\texttt{RFM\textsubscript{ab}}}.
\item \allchanges{We propose a centralized increment circuit that enables subarray-level parallelism, minimizing the counter update overhead during precharge.} 
\item We introduce a bank-level \allchanges{recovery} scheme that minimizes unnecessary stalls by isolating mitigation to the affected bank.
\item We integrate these two mechanisms into \papertitle{}, a low-overhead Rowhammer mitigation framework that improves performance while preserving the security guarantees of PRAC+ABO.
\item We evaluate \papertitle{} using Ramulator, showing lower performance overheads compared to PRAC+ABO with minimal hardware modifications.

\end{itemize}

\section{Background}
\sloppy

\begin{figure}
    \centering
    \includegraphics[width=1\linewidth]{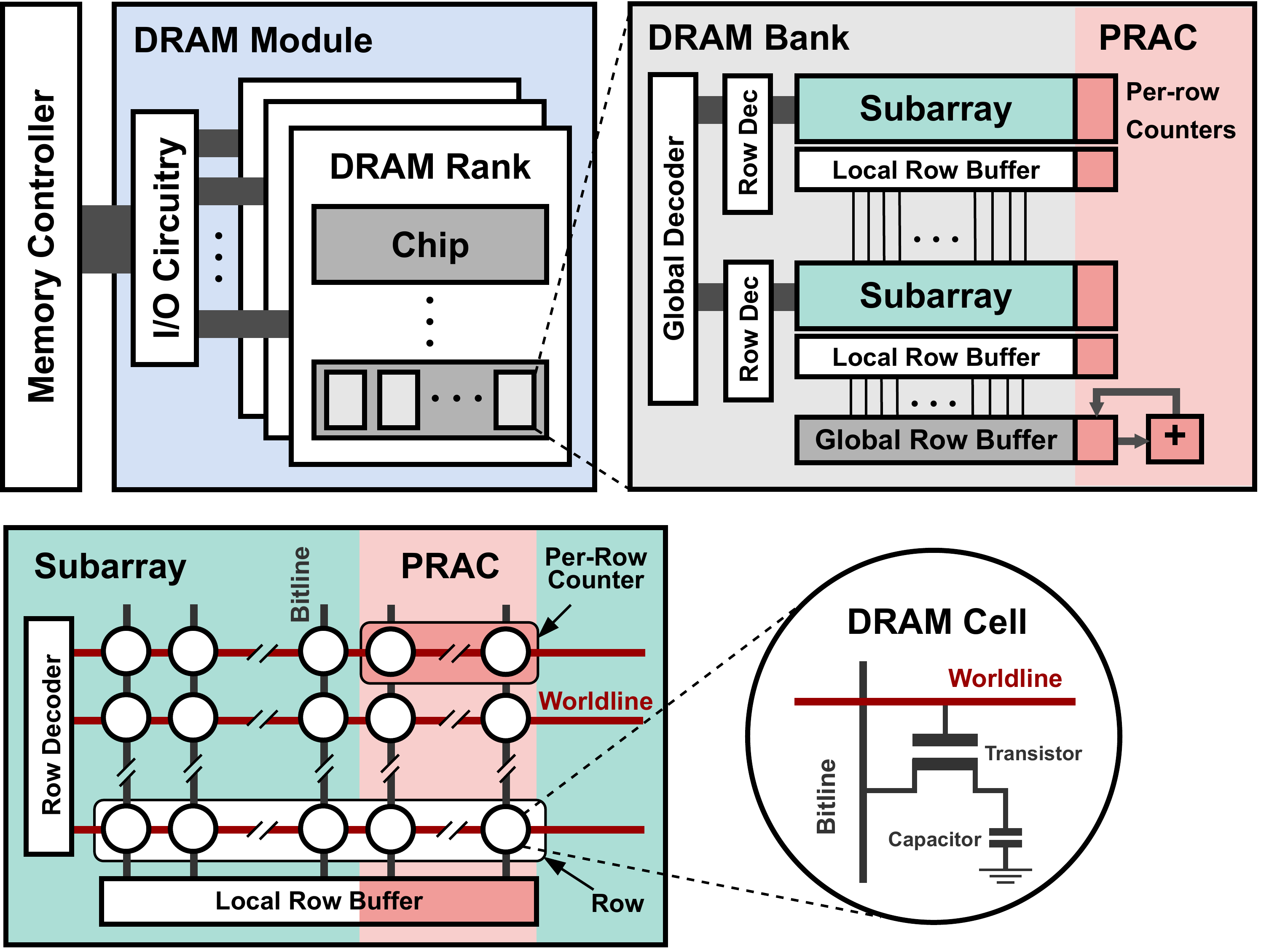}
    \caption{\allchanges{DRAM Organization and Per-Row Activation Counter (PRAC) extension.}}
    \label{fig:dram_arch}
    \vspace{-0.5cm}
\end{figure}

In this section, we provide an overview of key background topics, including DRAM architecture, the Rowhammer vulnerability, and existing mitigation strategies. We also describe the PRAC+ABO mechanism introduced in the DDR5 standard, which forms the basis for our proposed performance enhancements.

\subsection{DRAM Architecture and Parameters}

Modern DRAM is organized hierarchically into ranks, banks, subarrays, rows, and columns, as illustrated in Figure ~\ref{fig:dram_arch}. At the top of this hierarchy, the memory controller communicates with DRAM modules via a dedicated memory channel. Each module is composed of one or more ranks, which share access to the memory channel in a time-multiplexed fashion. A rank comprises several DRAM chips, and each chip contains multiple banks. Within each bank are numerous subarrays, forming the fundamental building blocks of memory storage. Each subarray is implemented as a two-dimensional array of cells, accessed via rows (wordlines) and columns (bitlines). A single DRAM cell includes a capacitor, which holds a bit of data as an electric charge, and an access transistor that enables read and write operations. Each subarray contains its own local row buffer, which temporarily holds the contents of an activated row. Only one subarray at a time can forward its data to the global row buffer shared across the bank, enabling read and write operations to proceed.
To access data, the memory controller issues an \textit{\textbf{Activate (ACT)}} command, which opens a specific row by transferring its contents into the row buffer. Subsequent read or write operations are performed on this open row, resulting in a low-latency row hit. If a different row within the same bank must be accessed, the current row must first be closed using a \textit{\textbf{Precharge (PRE)}} command, which restores the contents of the row buffer to the array and resets the bitlines.

DRAM access behavior is governed by a set of well-defined timing parameters specified by the JEDEC standard~\cite{jedec2024pressrelease}, as summarized in Table~\ref{tab:dram-timings}. \allchanges{We use DDR5-3200AN timings.} For example, the \textbf{Row Access Strobe (RAS)} latency defines the minimum delay between an ACT and a subsequent PRE command within the same bank. Additionally, to maintain data integrity, DRAM cells require periodic refreshing within a retention window denoted as \textbf{$t_{\mathrm{REFW}}$}, typically \emph{32ms}. To amortize the cost of refresh, DRAM divides its cells into 8192 refresh groups, and a \textbf{Refresh (REF)} command is issued every \textbf{$t_{\mathrm{REFI}}$} = \emph{3900ns} to refresh one group at a time.

\begin{table}[h]
    \centering
    \resizebox{0.47\textwidth}{!}{  
    \begin{tabular}{|>{\raggedright\arraybackslash}p{1.5cm}|>{\raggedright\arraybackslash}p{5.5cm}|>{\centering\arraybackslash}p{1.0cm}|>{\centering\arraybackslash}p{1.0cm}|}
        \hline
        \textbf{Parameter} & \textbf{Description} & \textbf{Base} & \textbf{PRAC} \\ \hline
        \( t_{RAS} \) & Minimum time a row must be kept open & 32ns & 16ns \\ \hline
        \( t_{RP} \) & Time to precharge an open row & \allchanges{15}ns & 36ns \\ \hline
        \( t_{RC} \) & Time between successive ACTs to a bank & \allchanges{47}ns & 52ns \\ \hline
        \allchanges{\( t_{RTP} \)} & \allchanges{Minimum time for a PRE
after a RD to the same bank} & \allchanges{7.5ns} & \allchanges{5ns} \\ \hline
\allchanges{\( t_{WR} \)} & \allchanges{minimum time for a PRE
after a WR to the same bank} & \allchanges{30ns} & \allchanges{10ns} \\ \hline

    \end{tabular}
    }
    \caption{\allchanges{DRAM Timing Standards}}
    \label{tab:dram-timings}
\end{table}

\subsection{Read Disturbance Attacks}

Modern DRAM is increasingly vulnerable to read disturbance effects, where the act of accessing one memory row can inadvertently influence the integrity of data stored in nearby rows. This phenomenon arises from the shrinking physical dimensions of DRAM cells and the reduced noise margins between them. As manufacturing processes continue to scale, the isolation between adjacent rows weakens, making DRAM cells more susceptible to charge leakage and electromagnetic coupling.

The most prominent example of a read disturbance attack is Rowhammer~\cite{kim2014flipping}, where an attacker repeatedly activates (i.e., opens) a single DRAM row, known as the aggressor row, within a refresh interval. If the number of activations exceeds a certain threshold, electrical interference can induce bit flips in adjacent victim rows. Since its discovery, Rowhammer has been shown to compromise system security in various ways, including enabling privilege escalation~\cite{seaborn2015exploiting}, breaking isolation between virtual machines~\cite{razavi2016flip}, and subverting browser-based sandboxes~\cite{gruss2016rowhammer}.

Beyond Rowhammer, newer variants continue to expose the fragility of DRAM. RowPress~\cite{luo2023rowpress} demonstrates that simply holding a row open for an extended duration—without repeated activations—can cause similar disturbance effects in neighboring rows. This shows that both temporal access frequency and row residency time can lead to data corruption, expanding the threat model beyond traditional Rowhammer.

The security implications of these attacks are severe. Bit flips in sensitive memory structures such as page tables, kernel space, or encryption keys can be exploited to gain arbitrary code execution, bypass memory isolation, or compromise confidentiality~\cite{frigo2020trrespass,gruss2018another,razavi2016flip,gruss2016rowhammer}.
To mitigate read disturbance attacks, a wide range of defenses have been proposed. These typically follow a two-phase approach:
(1) Detecting aggressor rows, using techniques such as memory controller-based row activation counters~\cite{kim2021hammerfilter}, probabilistic sampling~\cite{son2017making, kim2014flipping}, or in-DRAM tracking~\cite{jaleel2024pride, qureshi2024mint}; and (2) Applying preventive measures, such as Target Row Refresh (TRR)~\cite{marazzi2022protrr, frigo2020trrespass}, domain-aware memory allocation~\cite{saxena2024preventing}, or row remapping~\cite{saxena2024rubix}. While many of these defenses are effective at reducing bit flips, they often incur significant performance overheads and hardware complexity, especially when mitigation is applied conservatively to avoid false negatives. 
As DRAM continues to scale and disturbance thresholds fall, there is a growing need for efficient, low-latency, and fine-grained mitigation mechanisms that preserve performance without compromising reliability or security.

\subsection{PRAC and ABO in Modern DRAM Standards}

To address the growing threat of Rowhammer attacks, recent JEDEC standards have introduced two complementary in-DRAM mitigation mechanisms: \textbf{\textit{Per-Row Activation Counter (PRAC)}} and \textbf{\textit{Alert Back-Off (ABO)}}. These mechanisms aim to detect and mitigate malicious or excessive row activations efficiently, while remaining scalable for high-density DRAM systems.

\subsubsection{PRAC}

This feature is designed to mitigate Rowhammer attacks with minimal area and power overhead by embedding activation tracking directly within the DRAM array. \hlindigo{Specifically, PRAC extends each DRAM row with a dedicated activation counter.} Every time a row is \allchanges{precharged}, its corresponding counter is incremented by the increment logic in the global row buffer shared between all subarrays within a bank, as illustrated in Figure~\ref{fig:dram_arch}. This increment operation is performed during the precharge phase and is implemented as a \textit{read-modify-write (RMW)} sequence at the bank level. When a precharge command (\textit{PRE}) is issued, the DRAM bank reads the activation counter of the recently accessed row, updates its value, and writes it back before deactivating the wordline. This sequence introduces additional latency to the precharge operation, requiring DRAM timing parameters to be extended to accommodate the counter update overhead. While PRAC enables fine-grained tracking of row activations without relying on external memory controller storage, its bank-wide RMW implementation creates performance bottlenecks by delaying subsequent accesses to other non-conflicting subarrays until the counter update completes, as illustrated in top and middle rows of the Figure~\ref{fig:motiv} (a). The performance impact of this overhead is analyzed in detail in Section~\ref{sec:motivation}.

\subsubsection{ABO} 

\begin{figure}
    \centering
    \includegraphics[width=1.0\linewidth]{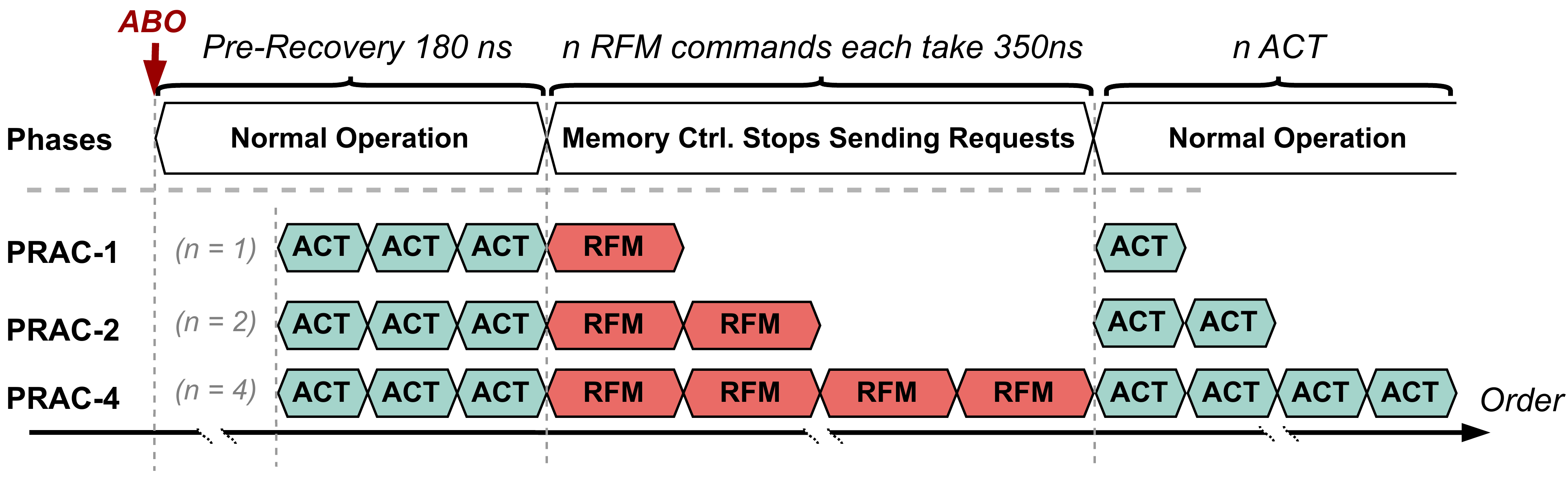}
    \vspace{-0.3in}
    \caption{\allchanges{Alert-back-off (ABO) overview}}
    \label{fig:abo_timing}
\end{figure}

This mechanism complements PRAC by providing a signaling interface between the DRAM device and the memory controller. When a PRAC counter in any bank reaches a pre-defined threshold, the corresponding bank sends an ABO signal to notify the memory controller that mitigation is needed. Upon receiving the ABO signal, the memory controller initiates a pre-recovery phase lasting \textit{180ns}, during which it continues to serve memory requests normally as shown in Figure ~\ref{fig:abo_timing}. \hlindigo{After this window, the controller issues an} \texttt{\hlindigo{RFM}\textsubscript{\hlindigo{ab}}} \hlindigo{command to trigger targeted mitigation}. Each \allchanges{\texttt{RFM\textsubscript{ab}}} command incurs a latency of \textit{350ns} and applies to all banks in the channel (typically 64 banks). During this period, the memory controller stalls requests to the \allchanges{channel} while \allchanges{the banks} perform recovery operations, such as refreshing \allchanges{victim} rows, as shown in the middle row of Figure~\ref{fig:motiv} (b). If multiple banks reach the threshold in close succession, the memory controller may need to issue multiple RFM commands, each separated by a recovery window. This can amplify the performance impact, especially under aggressive access patterns that frequently trigger mitigation events.

To prevent back-to-back ABO signals, the protocol requires a minimum number of \textit{ACT} commands between two consecutive alerts. In Figure~\ref{fig:abo_timing}, this value is denoted by $n$. It also corresponds to the number of required \textit{RFM} operations, as each \textit{RFM} can service only a limited number of victim rows per invocation. After all \textit{RFM} commands are completed, one additional \textit{ACT} is permitted. The allowed values for $n$ are 1, 2, or 4, as illustrated by \textit{PRAC-1}, \textit{PRAC-2}, and \textit{PRAC-4} in Figure~\ref{fig:abo_timing}, as well as in Figures in Sections~\ref{sec3} and~\ref{sec:eval}. Consequently, the minimum and maximum ABO intervals range from \allchanges{ \textit{350ns} to \textit{1500ns}}, each preceded by a fixed \textit{180ns} pre-recovery window.

\subsection{Mitigations Based On PRAC+ABO}

Several recent works build on the PRAC+ABO framework to strengthen its security guarantees and reduce its performance overheads. In this paper, we focus on three representative designs—Chronus~\cite{canpolat2025chronus}, QPRAC~\cite{woo2025qprac}, and MOAT~\cite{qureshi2024moat}—which propose distinct approaches to improving the effectiveness and efficiency of in-DRAM RowHammer mitigation.

\textbf{Chronus}~\cite{canpolat2025chronus} introduces architectural modifications to decouple PRAC counter updates from the critical path of DRAM access. By relocating per-row counters to a dedicated metadata subarray, Chronus reduces the serialization overhead associated with precharge operations. It also proposes enhancements to the ABO protocol by holding the alert signal until all mitigations are complete, ensuring stronger coordination between DRAM and the memory controller.

\textbf{QPARC}~\cite{woo2025qprac} revisits PRAC's security model as originally proposed in the Panopticon framework~\cite{bennett2021panopticon}, and identifies two new attack vectors that exploit timing gaps in counter updates. To address this, it introduces a priority-based queue that tracks frequently accessed rows and schedules them for mitigation through ABO. This queuing mechanism enables timely and targeted mitigation while preserving compatibility with the existing PRAC+ABO interface.

\textbf{MOAT}~\cite{qureshi2024moat} focuses on simplifying mitigation logic by replacing queue-based designs with a tracking mechanism. It uses two SRAM registers to monitor one hot row at a time, applying proactive refreshes below a configurable threshold and issuing ABO signals when the threshold is exceeded. While MOAT reduces hardware complexity, it may face limitations under workloads with multiple simultaneous hot rows.

\section{PRAC+ABO Limitations and Motivation}
\label{sec3}
\label{sec:motivation}

While PRAC+ABO provides a low-cost, JEDEC-standardized approach for in-DRAM Rowhammer mitigation, it introduces significant performance bottlenecks and security risks due to its coarse-grained design. This section presents a detailed analysis of the overheads introduced by PRAC updates and ABO-triggered channel-wide stalls. We also show how these issues can be exploited by adversaries to launch memory performance attacks. All insights are derived from cycle-accurate simulations, as described in Section~\ref{sec::experimental_method}.

\subsection{Impact of PRAC Updates on DRAM Latency}
\label{sec3.1}

PRAC extends each DRAM row with an activation counter that is incremented during the \texttt{Precharge} (\texttt{PRE}) command using a \textit{read-modify-write (RMW)} operation, as illustrated in Figure~\ref{fig:dram_arch}. This update is performed at the bank level, using logic typically located near the global row buffer. As a result, when a \texttt{PRE} command is issued, the DRAM must first read, modify, and write back the counter value associated with the activated row before deactivating the wordline and returning to an idle state. This \textit{serialization} delays subsequent memory operations, even when they target non-conflicting subarrays.
To accommodate this RMW sequence, DRAM timing parameters are modified. Specifically, the precharge latency increases from \allchanges{\textit{15ns}} to \textit{36ns}, and the activation-to-activation interval ($t_{RC}$) increases from \allchanges{\textit{47ns}} to \textit{52ns}. 
Interestingly, the row active time ($t_{RAS}$) is reduced from \textit{32ns} to \textit{16ns}, partially offsetting the impact. Nevertheless, the increase in $t_{RC}$ introduces a minimum \allchanges{\textit{5ns}} delay between successive row activations — particularly impactful in workloads with high \allchanges{row buffer} conflicts.

\begin{figure}
    \centering
    \includegraphics[width=1\linewidth]{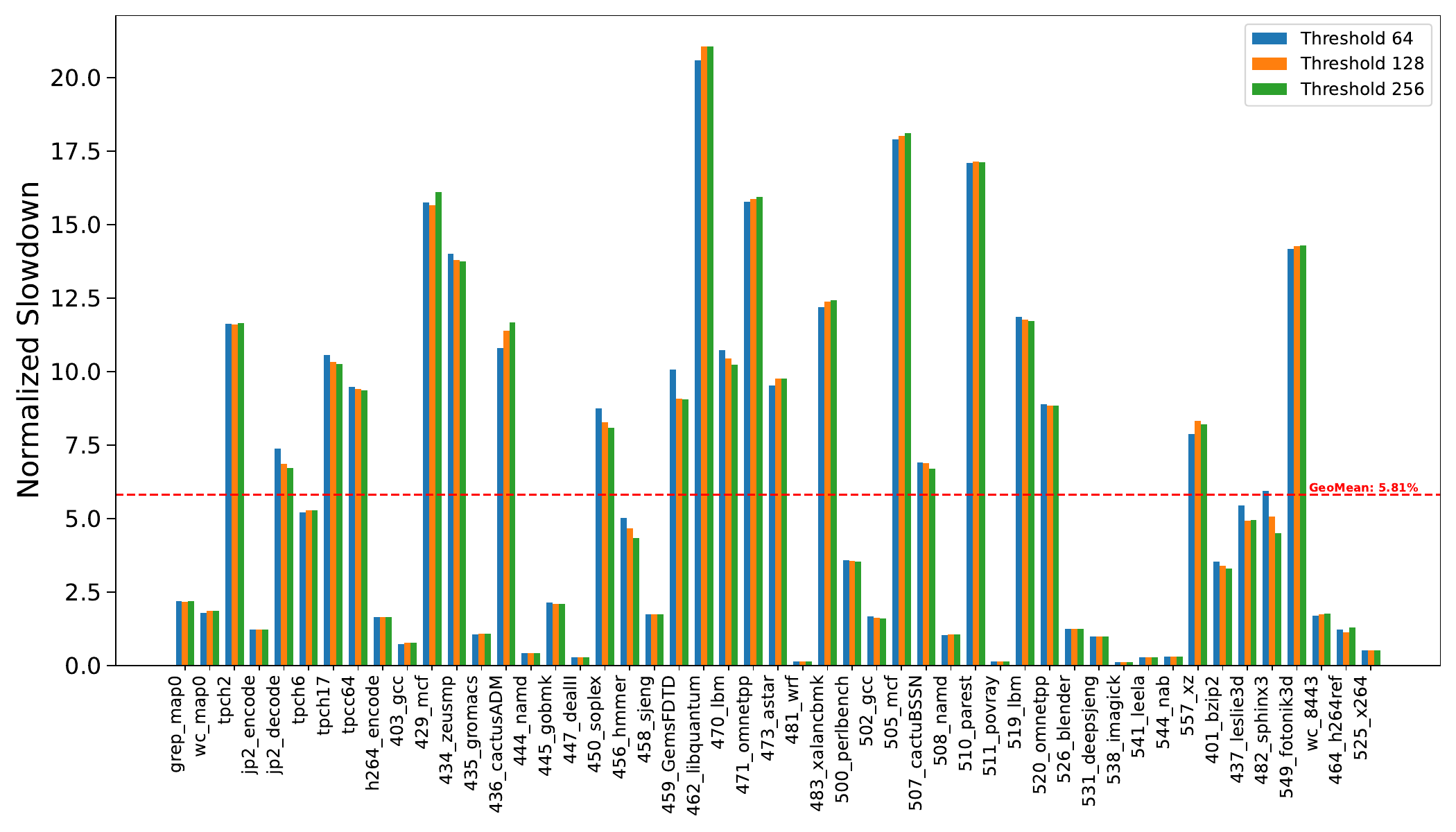}
    \caption{\allchanges{Percentage slow-down due to the increase in the latency of PRE command.}}
    \label{fig:motiv_slowdown}
\end{figure}

 To quantify this impact, we evaluated a set of memory-intensive workloads under two timing configurations: (1) a baseline DRAM configuration using standard DDR5 timings, and (2) a PRAC-enabled configuration with updated \allchanges{timings}, as specified in Table~\ref{tab:dram-timings}. The results, shown in Figure~\ref{fig:motiv_slowdown}, indicate that workloads experience an average slowdown of \allchanges{6\%, with a maximum of close to 20\% for \textit{462.libquantum}}. These slowdowns are primarily attributed to increased latency introduced by PRAC counter updates, particularly in scenarios involving frequent row activations within the same bank. \hlorange{Although the row access time increases by only 10\% (from 47ns to 52ns), the observed system-level performance overhead can reach up to 20\%. This discrepancy arises because rows are not always precharged immediately after access. In many cases, a row remains open for extended durations due to row buffer policy, reducing the impact of a longer row access time. However, the increased precharge latency, nearly a 100\% increase in the PRE timing, becomes the dominant contributor to performance degradation, especially when frequent rows are not immediately precharged.}

We also evaluated multiple Alert thresholds (64, 128, and 256) and observed only minor variations in slowdown across these configurations. This suggests that the performance degradation is largely due to static timing overheads from PRAC’s counter update logic, rather than the dynamic triggering of ABO or mitigation procedures. As expected, applications with low row conflict rates, such as \textit{447\_dealII} and \textit{444\_namd}, exhibited significantly less performance impact.

\begin{tcolorbox}[colback=gray!10,colframe=black,boxrule=0.5pt,arc=2mm,outer arc=2mm]
\textit{\textbf{Observation 1}: \allchanges{The updated DDR5 timings} introduce performance overhead, slowing down benign workloads by \allchanges{ geometric mean of 6}\%.}
\end{tcolorbox}

\begin{figure}
    \centering
    \includegraphics[width=1\linewidth]{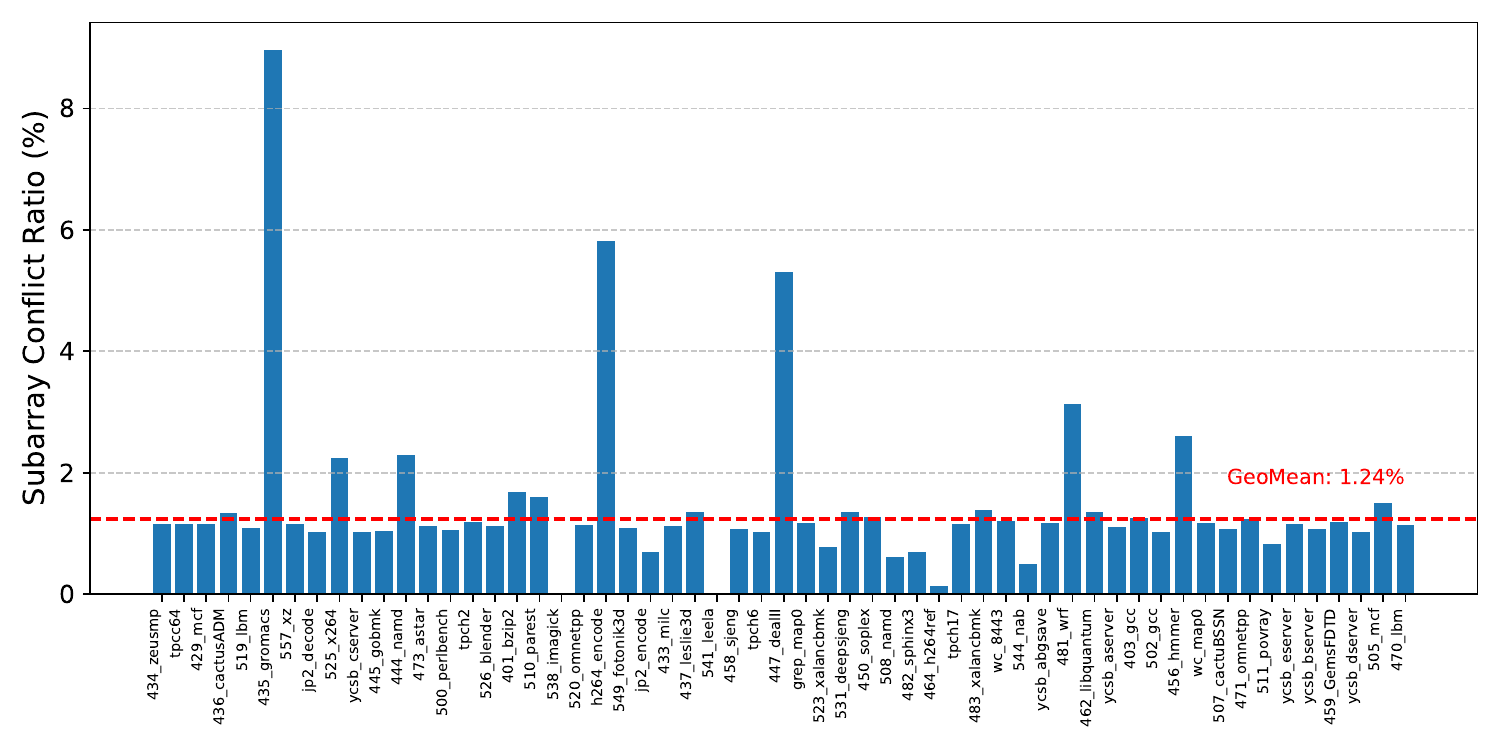}
    \vspace{-0.3in}
    \caption{\allchanges{Ratio(\%) of Subarray conflicts.}}
    \label{fig:subarray_conf}
\end{figure}

\hlorange{\papertitle{} proposes reducing the performance overhead of PRAC by performing counter updates within the local row buffers of individual subarrays. Modern DRAM architectures commonly adopt a density-optimized open-bitline structure}~\cite{dram_circuit_design, itoh_vlsi, chang2016low, luo2020clr}\hlorange{, which places sense amplifiers on both sides of a subarray and allows adjacent subarrays to share these amplifiers. As a result, when a subarray is performing a counter update, the memory controller must delay not only accesses to that subarray but also to its neighboring subarrays.}
\allchanges{To evaluate the practical impact of this constraint, we measure the percentage of subarray conflicts relative to total row buffer conflicts. Assuming a configuration with 256 subarrays per bank, Figure~\ref{fig:subarray_conf} presents the ratio of subarray conflicts, including those involving adjacent subarrays, to total row buffer conflicts. The results demonstrate that subarray conflicts are relatively rare, accounting for only 1.24\% on average across a diverse set of workloads.
This observation indicates that enabling PRAC updates at the subarray level can enhance performance by permitting concurrent accesses to subarrays that are not involved in counter updates.}



\subsection{Inefficiency of Channel-Wide Stall of \allchanges{\texttt{RFM\textsubscript{ab}}} }
\label{sec3.2}

The ABO mechanism notifies the memory controller when a row's activation count crosses the threshold. After the \textit{180ns} pre-recovery period, the controller issues an \allchanges{\texttt{RFM\textsubscript{ab}}} command, triggering a \textit{350ns} stall across the entire memory channel even if only a small subset of banks require mitigation.

\begin{figure}
    \centering
    \includegraphics[width=1\linewidth]{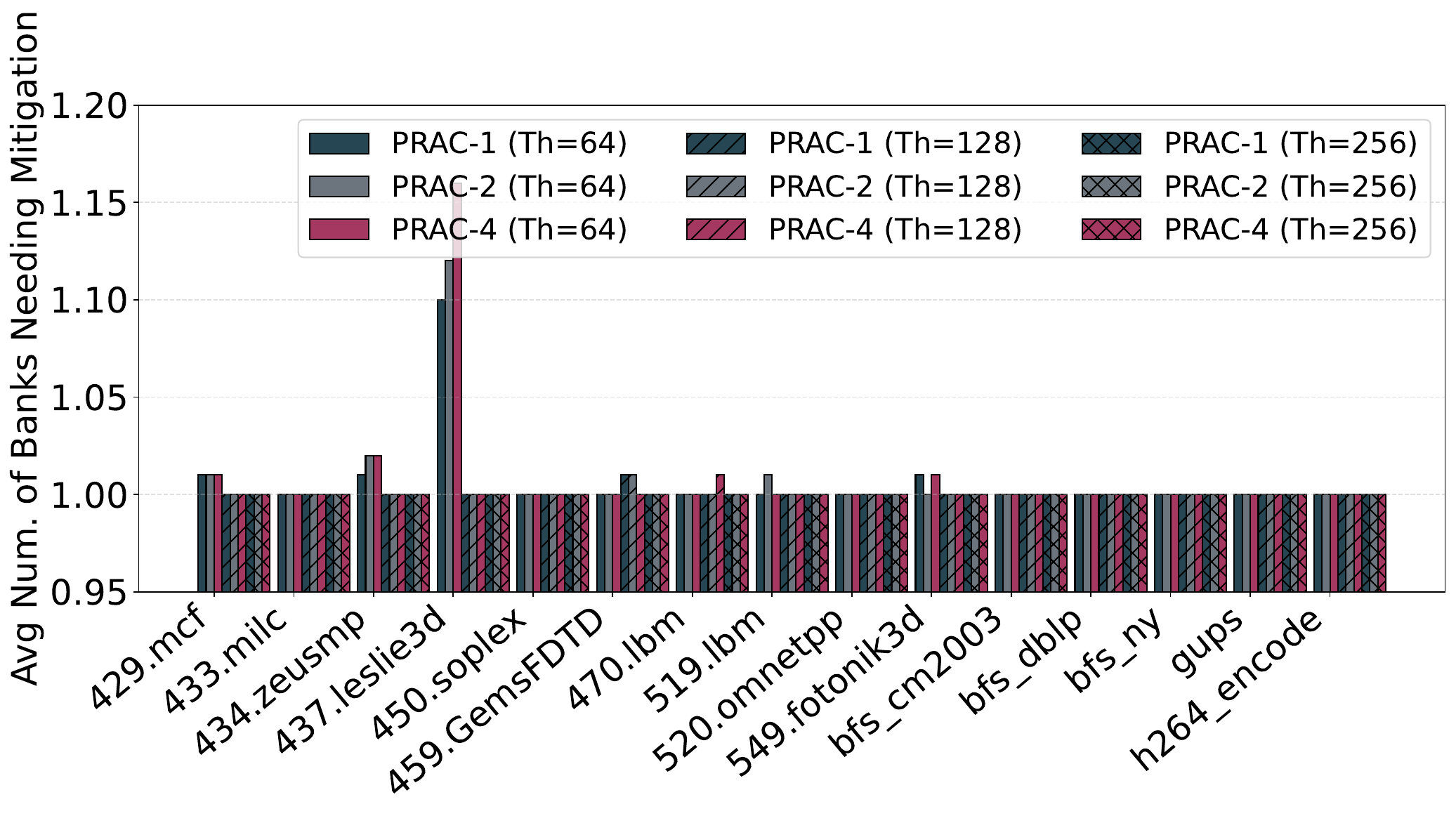}
    \caption{Average number of banks that need mitigation on an ABO signal for benign workloads}
    \label{fig:trueABO}
\end{figure}

\begin{figure}[ht]
\centering
\definecolor{color1}{HTML}{264653}
\definecolor{color1a}{HTML}{3e687a}
\definecolor{color1b}{HTML}{b1d5e5}
\definecolor{color2}{HTML}{6c757d}
\definecolor{color2a}{HTML}{adb5bd}
\definecolor{color2b}{HTML}{495057}
\definecolor{color3}{HTML}{a53860}
\definecolor{color3a}{HTML}{da7da0}
\definecolor{color3b}{HTML}{ab6e85}

\begin{tikzpicture}
    \begin{axis}[
        width = \linewidth,
        height = 3.2cm,
        major x tick style = transparent,
        ybar=0.7*\pgflinewidth,
        bar width=1.8pt,
        ymin=0,
        axis on top=false,
        ymax=4,
        ymajorgrids = true,
        enlarge x limits={0.04},
        xtick = data,
        ylabel = {\tiny Max Num. of Banks Needing Mitigation},
        xticklabel style={rotate=90,anchor=east,font=\tiny, xshift=0.2cm},
        yticklabel style={font=\tiny},
        legend cell align=left,
        legend style={
            at={(0.5,1.12)},
            anchor=south,
            legend columns=3,
            font=\tiny,
            column sep=0.1em,
            row sep=-0.1em,
            inner sep=1pt,
            draw=none,
            fill=white
        },
        clip=false,
        symbolic x coords={429.mcf, 433.milc, 434.zeusmp, 437.leslie3d, 450.soplex, 459.GemsFDTD, 470.lbm, 519.lbm, 520.omnetpp, 549.fotonik3d, bfs\_cm2003, bfs\_dblp, bfs\_ny, gups, h264\_encode},
        ylabel near ticks,
        ]
        
        \pgfplotstableread{
            Benchmark PRAC1 PRAC2 PRAC4
            429.mcf 2 2 2
            433.milc 1 1 1
            434.zeusmp 2 3 2
            437.leslie3d 4 3 3
            450.soplex 1 2 2
            459.GemsFDTD 2 1 1
            470.lbm 2 1 1
            519.lbm 2 2 1
            520.omnetpp 1 1 1
            549.fotonik3d 2 2 2
            bfs\_cm2003 1 1 1
            bfs\_dblp 1 1 1
            bfs\_ny 1 1 1
            gups 2 1 2
            h264\_encode 1 1 1
        }\datatableProc

        \addplot[draw=black, fill=color1, mark=none, bar shift=-6pt]
            table[x=Benchmark, y=PRAC1] {\datatableProc};
        \addplot[draw=black, fill=color2, mark=none, bar shift=-4pt]
            table[x=Benchmark, y=PRAC2] {\datatableProc};
        \addplot[draw=black, fill=color3, mark=none, bar shift=-2pt]
            table[x=Benchmark, y=PRAC4] {\datatableProc};

        \legend{PRAC-1 , PRAC-2 , PRAC-4}
    \end{axis}
\end{tikzpicture}
\caption{Maximum number of banks that need mitigation on an ABO signal for benign workloads.}
\label{fig:max_banks_mitigation}
\end{figure}
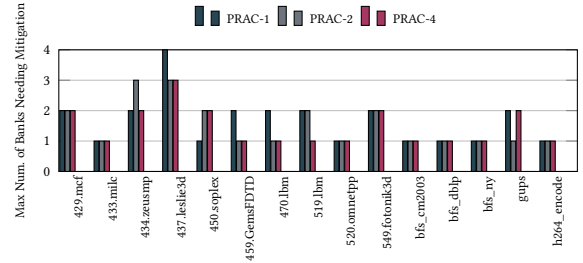

This coarse-grained stall mechanism is overly conservative. Across a wide range of benign workloads, our measurements show that, on average, only 1.16 out of 64 banks require mitigation when an ABO signal is raised (Figure~\ref{fig:trueABO}) and maximum of 4 banks need mitigation across all recovery periods (Figure~\ref{fig:max_banks_mitigation}). As a result, over 90\% of the banks are unnecessarily stalled during each recovery phase, significantly reducing memory-level parallelism and penalizing threads accessing unaffected banks.

\begin{figure}
    \vspace{-1cm}
    \centering
    \includegraphics[width=1\linewidth]{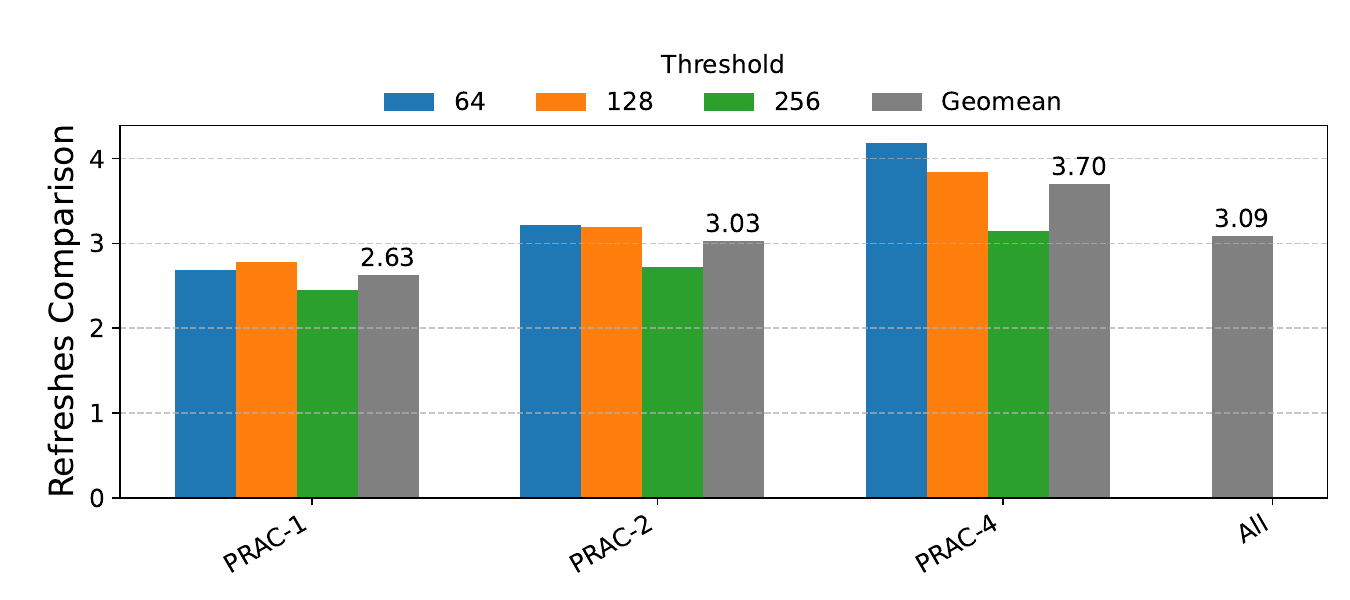}
    \caption{\allchanges{Comparison of number of RFM refreshes for opportunistic mitigation for PRAC-1, 2, and 4 and thresholds of 64, 128, and 256. }
    }
    \label{fig:num_refresh}

\end{figure}

\hlred{Although a bank may not require recovery, i.e., it does not contain any hot rows, the issuance of the} \texttt{RFM\textsubscript{ab}} \hlred{command blocks memory accesses to all banks within the channel. To utilize this otherwise idle period, prior works}~\cite{qureshi2024moat,canpolat2025chronus,woo2025qprac} 
\hlred{adopt \textit{opportunistic} mitigation strategies. These strategies proactively refresh potential victim rows whose counters are likely to reach the critical threshold in the near future. Specifically, upon issuance of }\texttt{RFM\textsubscript{ab}}\hlred{, each bank refreshes rows adjacent to the row with the highest activation count (i.e., the most likely aggressor), thereby ensuring that banks do not remain idle and reducing the need for issuing additional} \texttt{RFM\textsubscript{ab}} \hlred{commands in the future.

However, this approach introduces inefficiencies. Not all of the refreshes performed are strictly necessary, leading to redundant operations. To quantify this overhead, we compare the number of total} \texttt{RFM} \hlred{refreshes that are strictly required to those actually performed under opportunistic mitigation. As shown in Figure}~\ref{fig:num_refresh}\hlred{, the results, aggregated as the geometric mean across multiple PRAC variants (PRAC-1, PRAC-2, PRAC-4) and thresholds (64, 128, 256), demonstrate that opportunistic mitigation incurs more than a 3× increase in \texttt{RFM} refreshes. 
This analysis underscores a significant trade-off: while opportunistic mitigation reduces idle bank time and preempts future violations, it comes at the cost of approximately 70\% higher \texttt{RFM}-related energy consumption in DRAM.

}

\begin{tcolorbox}[colback=gray!10,colframe=black,boxrule=0.5pt,arc=2mm,outer arc=2mm,]
\textit{\textbf{Observation 2}: In benign workloads, fewer than 10\% of banks require mitigation during an ABO-triggered recovery period, \allchanges{while opportunistic mitigation performs 3x more recovery refreshes than needed.}}
\end{tcolorbox}

\subsection{Exploiting ABO: Performance Attacks} 
\label{sec3.3}

The coarse-grained nature of the ABO protocol introduces a new form of unfairness in DRAM systems, making them susceptible to performance \allchanges{slowdown} attacks. Because the ABO signal triggers channel-wide stalls without identifying the specific bank responsible for the excessive activation, a malicious actor can exploit this limitation to repeatedly disrupt system-level memory access.

In particular, an attacker can intentionally issue frequent row activations to a single bank to induce an ABO event. Since the memory controller stalls the entire memory channel upon receiving an alert, without visibility into which bank requires mitigation, benign workloads distributed across other banks also suffer from the resulting stall. This lack of spatial granularity makes it extremely difficult to attribute the cause of the alert or to selectively suppress malicious access patterns, rendering existing defenses ineffective against such attacks.
\hlred{
MOAT}~\cite{qureshi2024moat} \hlred{proposed the Torrent-of-Staggered-ALERT (TSA) attack, a performance degradation strategy that carefully coordinates ALERTs across multiple DRAM banks. In this attack, each bank repeatedly activates a small set of rows (e.g., rows A, B, C, D, E) to trigger an ALERT. Crucially, banks issue ACTs in a staggered fashion: a bank only initiates activation once all targeted rows in the previous bank have caused ALERTs and entered the mitigation phase. This serialized activation ensures that when one bank is under mitigation, other banks are forced to stall, as there are no eligible rows to activate without interference. This deliberate serialization of ALERTs forms a torrent of staggered mitigation events, significantly throttling memory concurrency. The attack was shown to reduce system throughput by 24\% with four banks and up to 52\% with 17 banks, aligning with the tFAW constraint.}

\begin{figure}
    \centering
    \includegraphics[width=1\linewidth]{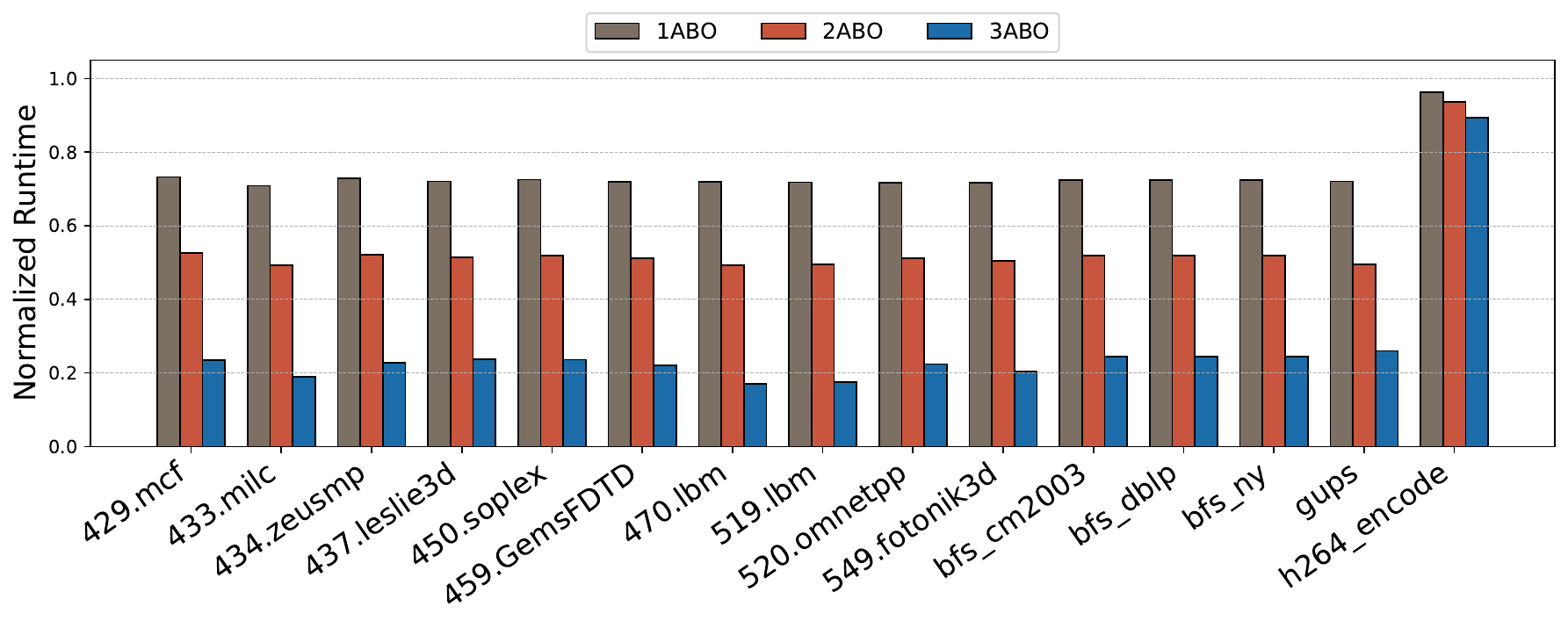}
    \caption{{Percentage slowdown of benign applications when there is a 1, 2 and 3 ABO alerts happening in each \texttt{tREFI}}}
    \label{fig:slowdownABO}
\end{figure}

\allchanges{BreakHammer~\cite{canpolat2024breakhammer} introduces a score-based mitigation framework that assigns a score to each hardware thread based on its contribution to Rowhammer mitigation events. This approach is particularly effective when mitigation logic is implemented within the MC, where score attribution can be performed at finer spatial granularity, such as individual banks or row activations. However, performance degradation attacks may evade detection under BreakHammer when mitigations are signaled via coarse-grained mechanisms such as the Alert Back-Off (ABO) signal. 

A broader class of memory performance degradation attacks exploits the observation that Rowhammer mitigations, such as recovery operations, are triggered more frequently at lower thresholds. As the mitigation frequency increases, overall system performance degrades significantly. To evaluate the performance cost associated with frequent mitigation, we study the impact of varying thresholds— 64, 32, and 16 as considered in prior works~\cite{woo2025qprac,canpolat2025chronus}.
}
\hlpurple{Assuming a refresh interval (\texttt{tREFI}) of 3900 ns and a refresh operation latency (\texttt{REF}) of 410 ns, approximately 67 row activations can be issued within one \texttt{tREFI} interval. We consider a closed-row memory policy where each activated row is closed after access. For a realistic scenario, an attacker can alternate between two rows to issue repeated activations. Under a threshold of 64, this behavior can typically trigger one Alert per \texttt{tREFI}. For thresholds of 32 and 16, the number of ALERTs increases proportionally, allowing an attacker to induce two to three (or even four) ALERTs within the same interval. Based on this analysis, we evaluate performance degradation under scenarios involving one, two, and three ALERTs per \texttt{tREFI} duration to quantify the impact of aggressive mitigation triggering.
}
As shown in Figure~\ref{fig:slowdownABO}, even a single ABO event per interval results in a 20–30\% slowdown for most applications. With three alerts per interval, performance degradation exceeds 80\% in several cases. Interestingly, workloads such as \texttt{h264\_encode} exhibit resilience due to their lower sensitivity to DRAM stalls, but most others are highly vulnerable.
These results demonstrate that the lack of bank-level precision in ABO signaling can be exploited by attackers to inflict disproportionate slowdowns on benign threads, emphasizing the need for spatially-aware mitigation mechanisms.

\begin{tcolorbox}[colback=gray!10,colframe=black,boxrule=0.5pt,arc=2mm,outer arc=2mm,]
\textit{\textbf{Observation 3}: Coarse-grained ABO signaling enables attackers to trigger repeated channel-wide stalls, leading to excessive performance loss.} 
\end{tcolorbox}

\section{\papertitle{} Design and Implementation}
To overcome the limitations of PRAC+ABO, we introduce \papertitle{}—an enhanced version that builds upon the original framework. \papertitle{} extends the underlying mechanisms to significantly reduce performance overheads while maintaining the robust security guarantees offered by state-of-the-art Rowhammer mitigation techniques. This section presents an overview of the design and implementation of \papertitle{}.

\subsection{Overview and Design Goals}
The core design principle of \papertitle{} is to minimize the performance overheads inherent in the original PRAC+ABO mechanism. Specifically, the design aims to achieve the following two objectives:

\begin{enumerate}[label={{\bfseries G\arabic*}:},leftmargin=2em]
\item \textbf{Reduce PRAC update latency through subarray-level decoupling.} \allchanges{Enable subarray-level counter updates to allow overlapping the counter update time of a row with activation of the next row when rows are non-conflicting at subarray-level, eliminating the \allchanges{\emph{21ns}} counter update delay.} 
\item \allchanges{\textbf{Minimize unnecessary memory stalls with fine-grained recovery command.} Refine \texttt{RFM\textsubscript{ab}} to operate at bank-level granularity, allowing the memory controller to stall only the affected banks instead of the entire channel.}
\end{enumerate}



\subsection{Hardware Modifications}

\papertitle{} introduces minimal hardware modifications to enable fine-grained, low-latency RowHammer mitigation by enhancing the PRAC+ABO feature.

\allchanges{ 
First, to support subarray-level PRAC updates, the traditional bank-level increment logic, typically located near the global row buffer, is replaced with a centralized increment circuit that is connected to local row buffers in subarrays through a different bus (8-wire bus for 8-bit counters). This design allows counter updates without delaying the subsequent memory accesses to the memory bank that do not conflict at the subarray level}. Correspondingly, the memory controller is extended with subarray mapping logic and an address decoder capable of identifying the target subarray for each memory request.

Second, to enable fine-grained bank-level recovery stalling, the DRAM chip is modified to include a new control register called the Bank Alert (BA) register. This register contains one bit per bank, where each bit indicates whether a given bank has any rows with an activation number higher than the \allchanges{Alert} threshold. The contents of the BA register are communicated to the memory controller and serve as a mask, allowing the controller to selectively stall only the \allchanges{requests to the} banks under mitigation while continuing to issue commands to unaffected banks.



\subsection{\papertitle{} Mechanism}

\papertitle{} introduces two core mechanisms to reduce the performance overheads associated with the PRAC+ABO framework while maintaining its security guarantees against RowHammer attacks.

\subsubsection{Subarray-Level PRAC Updates} 

\begin{figure}
    \centering
    \includegraphics[width=0.8\linewidth]{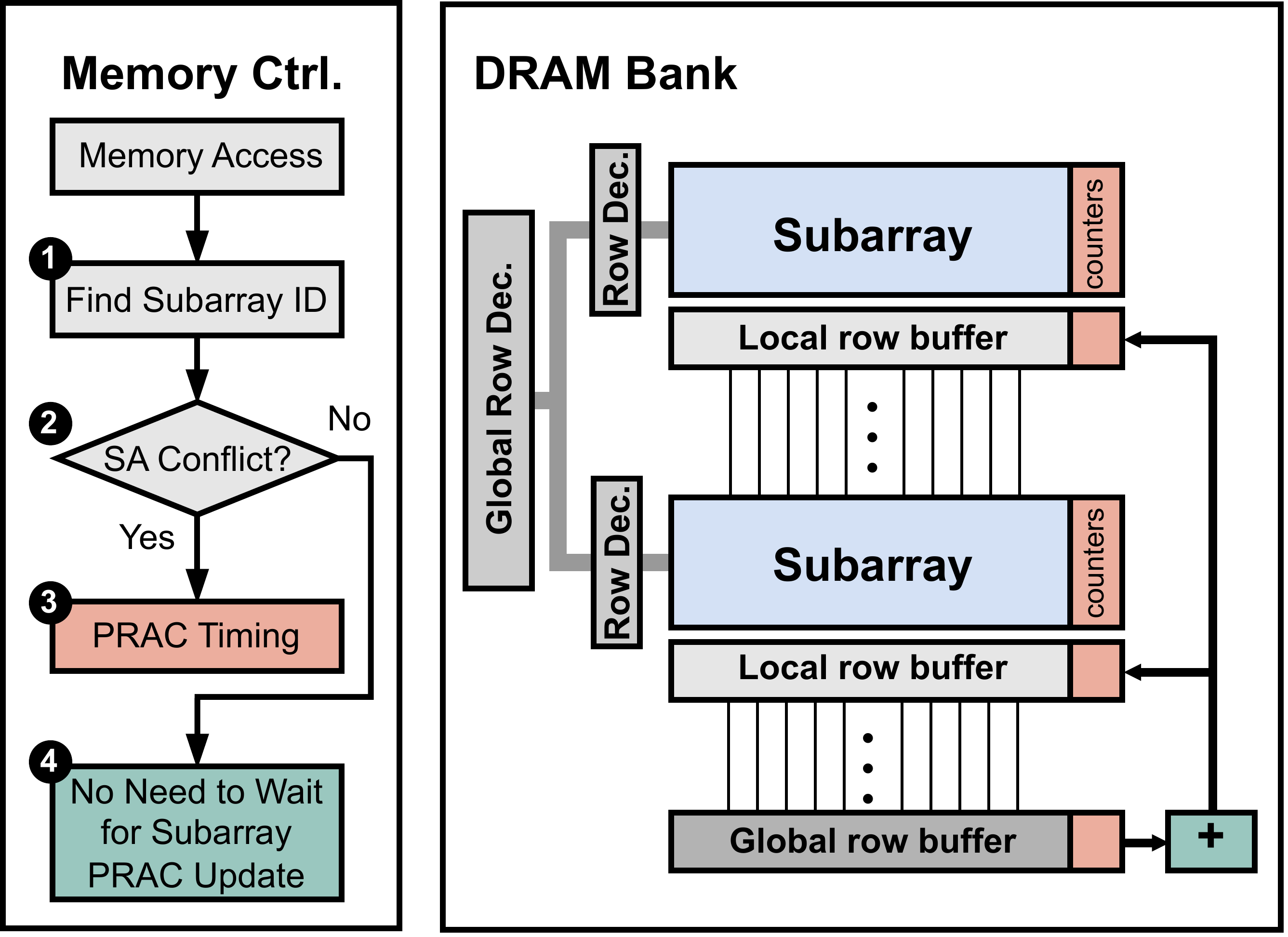}
    \caption{\allchanges{\papertitle{} adds subarray-level PRAC increment logic to enable independent counter updates and avoid bank-wide stalls across all subarrays.}}
    \label{fig:subarray_mechanism}
\end{figure}

This mechanism improves subarray-level parallelism by enabling PRAC counter updates at the subarray level, rather than enforcing bank-wide delays. In the original PRAC design, counter updates introduce additional latency, particularly during precharge operations, because the increment logic is shared at the bank level. This design forces all subarrays within the bank to stall while the update completes. \allchanges{ \papertitle{} addresses this limitation by introducing a centralized PRAC increment logic that connects to the local row buffers of all subarrays
, as illustrated in Figure~\ref{fig:subarray_mechanism}.
}

\begin{figure}
    \centering
    \includegraphics[width=0.8\linewidth]{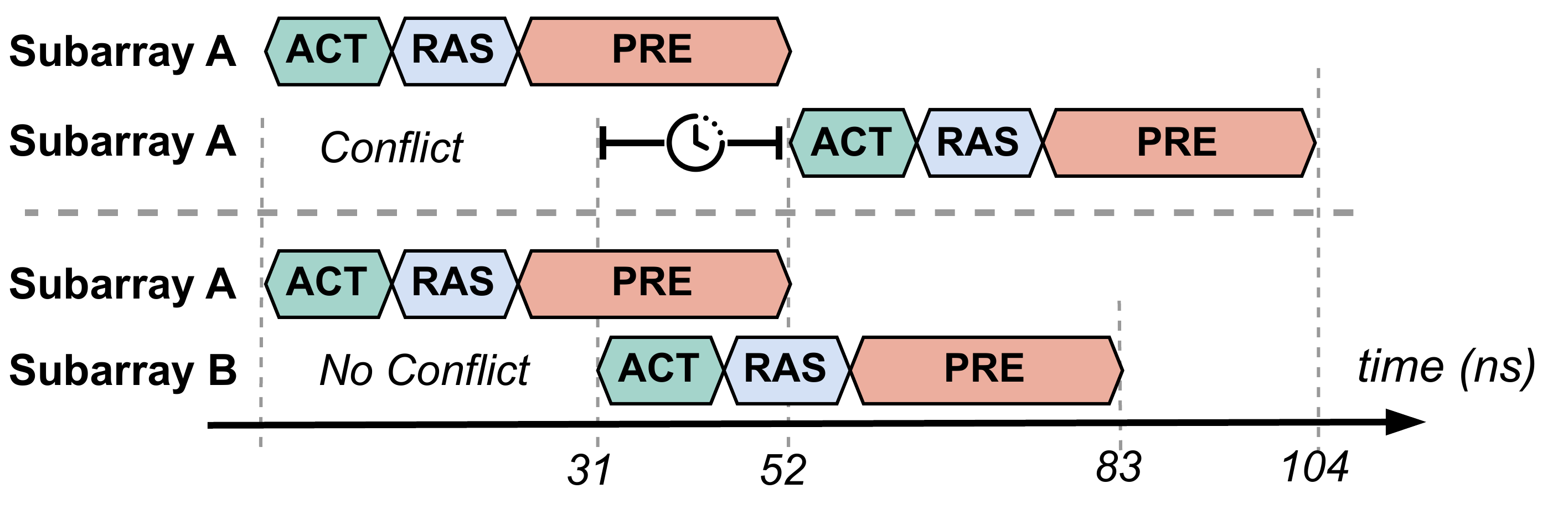}
    \caption{\allchanges{Illustration of conflicting (top row) and non-conflicting (bottom row) memory accesses across subarrays, demonstrating how \papertitle{} uses parallelism by avoiding unnecessary delay.}}
    \label{fig:example_subarray}
    \vspace{-0.5cm}
\end{figure}

Upon receiving a memory access request, the memory controller computes the subarray identifier (Subarray ID) for the target row \circled{1}. If the new request targets \allchanges{the same subarray or nearby subarray} that is currently undergoing a PRAC update (i.e. subarray conflict) \circled{2}, the controller applies the required timing constraints to ensure that the update completes before issuing a new activation \circled{3}. In contrast, if the access targets a different \allchanges{non-conflicting} subarray, the controller can proceed with the activation command immediately, without waiting for the PRAC update to finish \circled{4}. \allchanges{In this case, counter value of the precharged row is forwarded to the increment unit and the entire row is transferred from the global to the local row buffer. This precharge operation takes 15 ns to complete. After this delay, the increment circuit updates the counter and transmits the new value to the corresponding local row buffer via a dedicated counter data bus.
} This approach allows subarrays to operate independently, preventing unrelated memory accesses from being delayed by PRAC updates in other subarrays, as illustrated in Figure~\ref{fig:example_subarray}. As a result, \papertitle{} reduces access latency, particularly in workloads that exhibit diverse subarray access patterns. 

\subsubsection{\allchanges{Bank-Level Stall for Recovery}}

\begin{figure}
    \centering
    \includegraphics[width=0.8\linewidth]{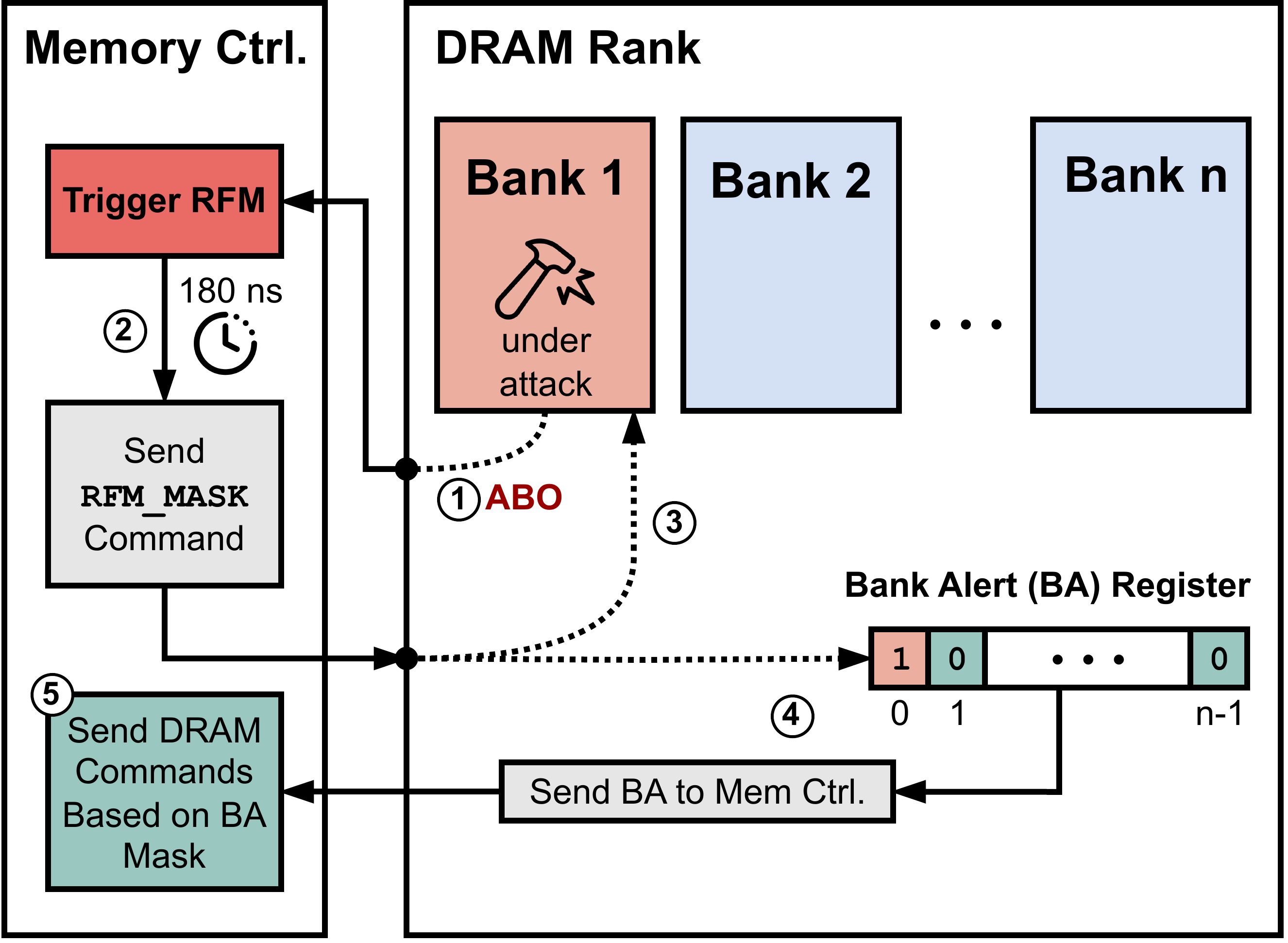}
    \caption{Bank-Level \allchanges{Recovery Stall} in \papertitle{} Using the Bank Alert (BA) Register and RFM\_MASK.}
    \label{fig:bank_mechanism}
    \vspace{-5pt}
\end{figure}

\papertitle{} addresses the limitation of coarse-grained mitigation in the original PRAC+ABO framework by introducing bank-level \allchanges{recovery stall}, enabling the memory controller to restrict mitigation only to the affected banks rather than stalling the entire memory channel. This selective mitigation is made possible through the use of a BA register, which functions as a mask to identify and isolate banks requiring Rowhammer \allchanges{mitigation} as illuatrated in Figure~\ref{fig:bank_mechanism}.

When the PRAC counter of a bank exceeds the Alert threshold, the DRAM device sends the ABO signal to the memory controller \wcircled{1}. Upon receiving this signal, the memory controller enters a pre-recovery window, lasting 180 nanoseconds, during which it continues to process memory requests as normal \wcircled{2}. After this window, the controller issues \allchanges{\texttt{RFM\textsubscript{ab}}} command to the DRAM to initiate recovery. \papertitle{} repurposes this command into a new command, termed \verb|RFM_MASK| (\allchanges{Masked Refresh Management}), which performs two key functions\allchanges{; it initiates recovery and functions as a register read command.}
First, the \verb|RFM_MASK| command triggers the in-DRAM recovery mechanism for the affected bank \wcircled{3}. Second, it returns the contents of the BA register to the memory controller, indicating which banks are currently under mitigation \wcircled{4}. \hlgray{The BA register is reset once it is read. After its reset, a bank can set the corresponding bit in the register.} The controller uses this information as a bitmask to manage access granularity during recovery. For any new memory request, the memory controller checks whether the target bank is marked as active in the BA register. If the request targets an unaffected bank, the controller proceeds to issue the command without delay. However, if the request targets a bank under recovery, the controller stalls the request until the mitigation phase completes \wcircled{5}. 

\begin{figure}
    \centering
    \includegraphics[width=1\linewidth]{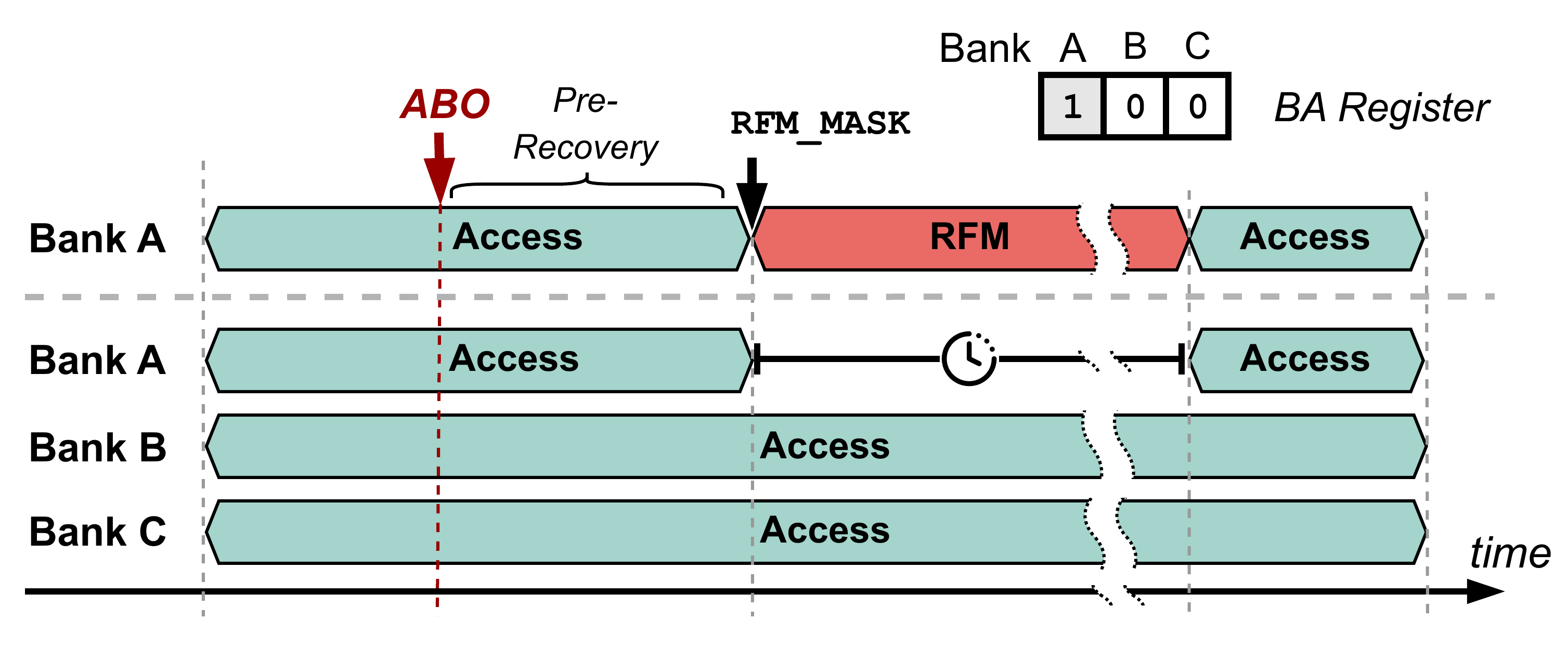}
    \caption{In this example, Bank 1 triggers ABO and enters recovery. Based on the BA mask, accesses to Banks 1 are stalled, while Banks 2 and 3 remain accessible.}
    \label{fig:example_bank}
\end{figure}

Figure~\ref{fig:example_bank} illustrates how \papertitle{} leverages bank-level \allchanges{recovery stall} to improve memory parallelism during RFM \allchanges{refreshes}. In this example, Bank 1 exceeds the \allchanges{Alert} threshold and triggers an ABO signal, initiating a short pre-recovery phase followed by the recovery phase. \hlgray{The banks modify their own bits in the register. Bank 1 will set its bit to 1.} Upon receiving the \verb|RFM_MASK| command the DRAM sends the contents of the BA register (\verb|0b1001|) to the MC, which encodes the mitigation status of all banks. The BA mask in this case indicates that Bank 1 is under mitigation. As a result, the memory controller selectively stalls requests to Bank 1 while continuing to issue accesses to unaffected Banks 2 and 3. This fine-grained handling contrasts with the baseline PRAC+ABO design, where all banks would be stalled upon any ABO event. \papertitle{} thus reduces unnecessary interference and improves memory-level concurrency during mitigation.

\section{Experimental Methodology}
\label{sec::experimental_method}



We utilize a cycle-accurate, open-source memory simulator, Ramulator 2.0~\cite{luo2023ramulator,ramulator2}. Our evaluation is on the existing PRAC+ABO framework and our proposed solution, \papertitle{}. We use the provided PRAC+ABO with RFM in Ramulator2.0. Furthermore, we evaluate Chronus~\cite{canpolat2025chronus} with \papertitle{} and compare the performance.

\begin{table}[b]
\centering
\caption{Simulator Configurations}
\label{tab::exp_eval}
\resizebox{\linewidth}{!}{
\renewcommand{\arraystretch}{0.9} 
\begin{tabular}{ll}
\hline
\addlinespace[0.2em] 
\textbf{Parameter} & \textbf{Configuration}               \\\hline
\addlinespace[0.2em] 
CPU                        & 4-core, 4.2\,GHz, 128-entry instruction window \\
Last-Level Cache           & 2MB per core, total 8MB (16-way set-associative) \\
Memory Controller          & 32-entry read/write queues \\
                           & Scheduling: FR-FCFS + Cap of 4\\
                           & Address mapping: MOP \\
Main Memory                & DDR5 DRAM, 1 channel, 2 ranks, 8 bank groups, \\
                           & 4 banks/group, 64K rows/bank \\\hline
\end{tabular}%
}
\end{table}
System configuration is presented in table~\ref{tab::exp_eval}. We use a pool of traces from SPEC CPU2006~\cite{spec2006}, SPEC CPU2017~\cite{spec2017}, TPC~\cite{tpc}, MediaBench~\cite{fritts2009mediabench} and YCSB~\cite{cooper2010benchmarking}.  The category of each trace is determined based on row buffer conflict per kilo instructions (RBMPKI). \allchanges{The traces are divided into High (H: $\geq$10 RBMPKI), Medium (M: 2--10 RBMPKI), and Low (L: $<$2 RBMPKI) memory usage, as shown in Table~\ref{tab:memory_usage}.}  For evaluation, we use 4 different traces to form groups of HHHH, MMMM, LLLL, HHMM, MMLL, and LLHH as mixed workloads, represented in Table~\ref{tab:workload_mapping}. For evaluation, we use \allchanges{10} workloads from each group and run 100M instructions for simulation. \hlpurple{Our simulation configuration aligns with previous  wroks}~\cite{olgun2024abacus,canpolat2025chronus,woo2025qprac}.
\begin{figure*}
    \centering
    \includegraphics[width=1\linewidth]{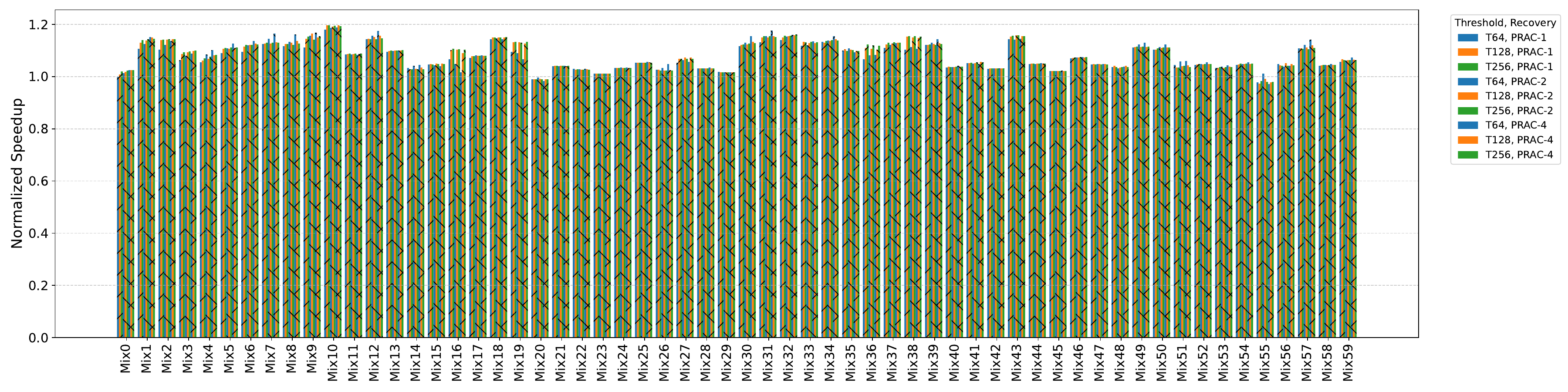}
    \caption{\allchanges{Performance evaluation of \papertitle{} using normalized speedup over PRAC+ABO}}
    \label{fig:opp_prac_grace}
    \vspace{-0.5cm}
\end{figure*}

\begin{table}[htbp]
    \centering
    \scriptsize 
    \resizebox{\linewidth}{!}{%
    \begin{tabular}{p{3em}p{\dimexpr\linewidth-8em\relax}}
        \hline
        \addlinespace[0.2em]
        \textbf{Memory Usage} & \textbf{Benchmarks} \\
        \hline
        \addlinespace[0.1em]
        High (H) & 429.mcf, 433.milc, 434.zeusmp, 450.soplex, 459.GemsFDTD, 462.libquantum, 470.lbm, 482.sphinx3, 483.xalancbmk, 510.parest, 519.lbm, 549.fotonik3d, \allchanges{gups, 520.omnetpp} \\
        \hline
        Medium (M) & 436.cactusADM, 473.astar, 507.cactuBSSN, 557.xz, jp2\_decode, jp2\_encode, tpcc64, tpch17, tpch2, wc\_8443, wc\_map0, ycsb\_aserver, ycsb\_bserver, ycsb\_cserver, ycsb\_eserver \\
        \hline
        Low (L) & 401.bzip2, 403.gcc, 435.gromacs, 444.namd, 445.gobmk, 447.dealII, 456.hmmer, 458.sjeng, 481.wrf, 500.perlbench, 502.gcc, 508.namd, 511.povray, 523.xalancbmk, 526.blender, 538.imagick, 541.leela, 544.nab, grep\_map0, tpch6, ycsb\_abgsave \\
        \hline
    \end{tabular}%
    }
    \caption{\allchanges{Grouping of Benchmarks by Memory Usage}}
    \label{tab:memory_usage}
\end{table}

\begin{table}[htbp]
    \centering
    \small 
    \begin{tabular}{ll}
        \hline
        \addlinespace[0.2em]
        \textbf{Workload} & \textbf{Mixed Types} \\
        \hline
        \addlinespace[0.2em]
        HHHH & Mix0 to 9 \\
        MMMM & Mix10 to 19 \\
        LLLL & Mix20 to 29 \\
        HHMM & Mix30 to 39 \\
        MMLL & Mix40 to 49 \\
        LLHH & Mix50 to 59 \\
        \hline
    \end{tabular}
    \caption{\allchanges{Mapping of Workload Types to Mixed Types}}
    \label{tab:workload_mapping}
\end{table}

\section{Evaluation}

\label{sec:eval}
\allchanges{

In this section we evaluate the performance improvements and compare results against similar state-of-the-art works. First, \papertitle{} is compared against standard PRAC with an Alert Back-Off signal to show performance benefit over standard JEDEC specifications. Then, we evaluate \papertitle{}'s effectiveness on one PRAC+ABO-based solution, QPRAC. We also show how resilient \papertitle{} is to chain attacks~\ref{sec3.3}.}
\subsection{Performance and Energy Evaluation}

\begin{figure}
    \centering
    \includegraphics[width=1\linewidth]{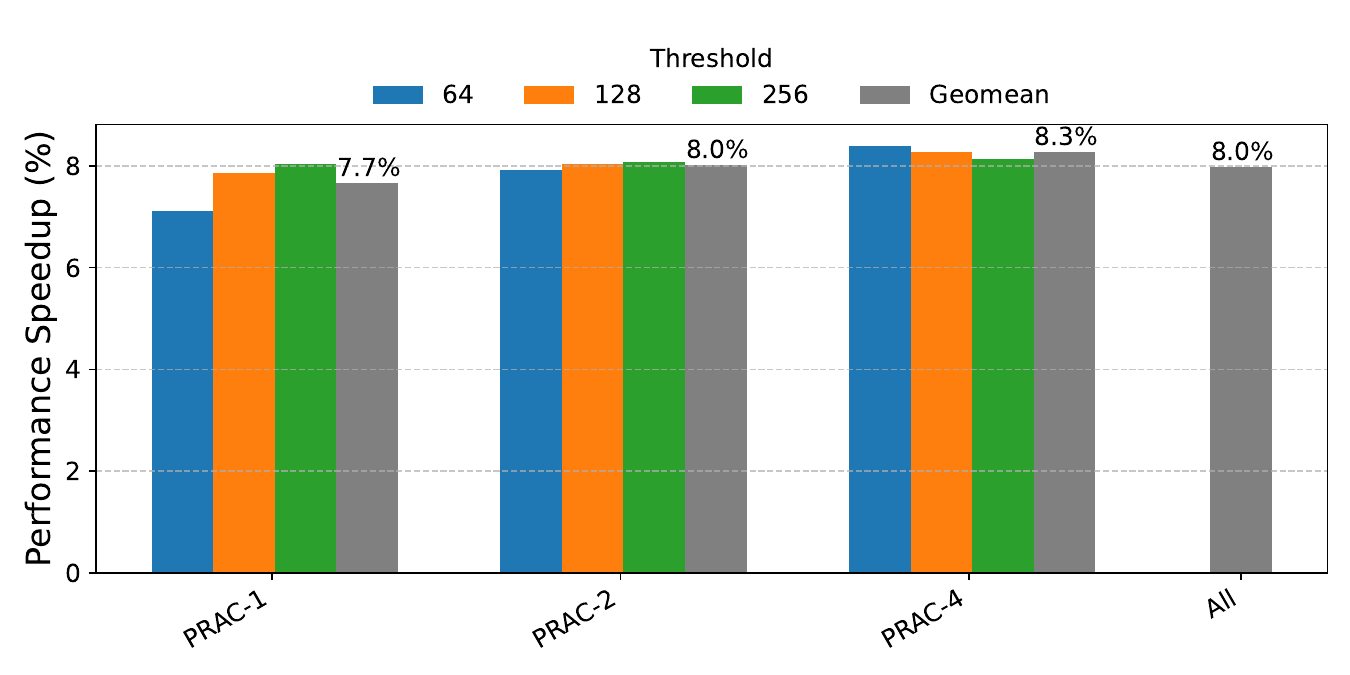}
    \vspace{-20pt}
    \caption{\allchanges{Performance Comparison of \papertitle{} and opportunistic PRAC+ABO for recoveries of 1, 2, and 4 and thresholds of 64, 128, and 256}}
    \label{fig:perf_eval_grace_prac}
    \vspace{-0.5cm}
\end{figure}

\begin{figure}
    
    \centering
    \includegraphics[width=0.8\linewidth]{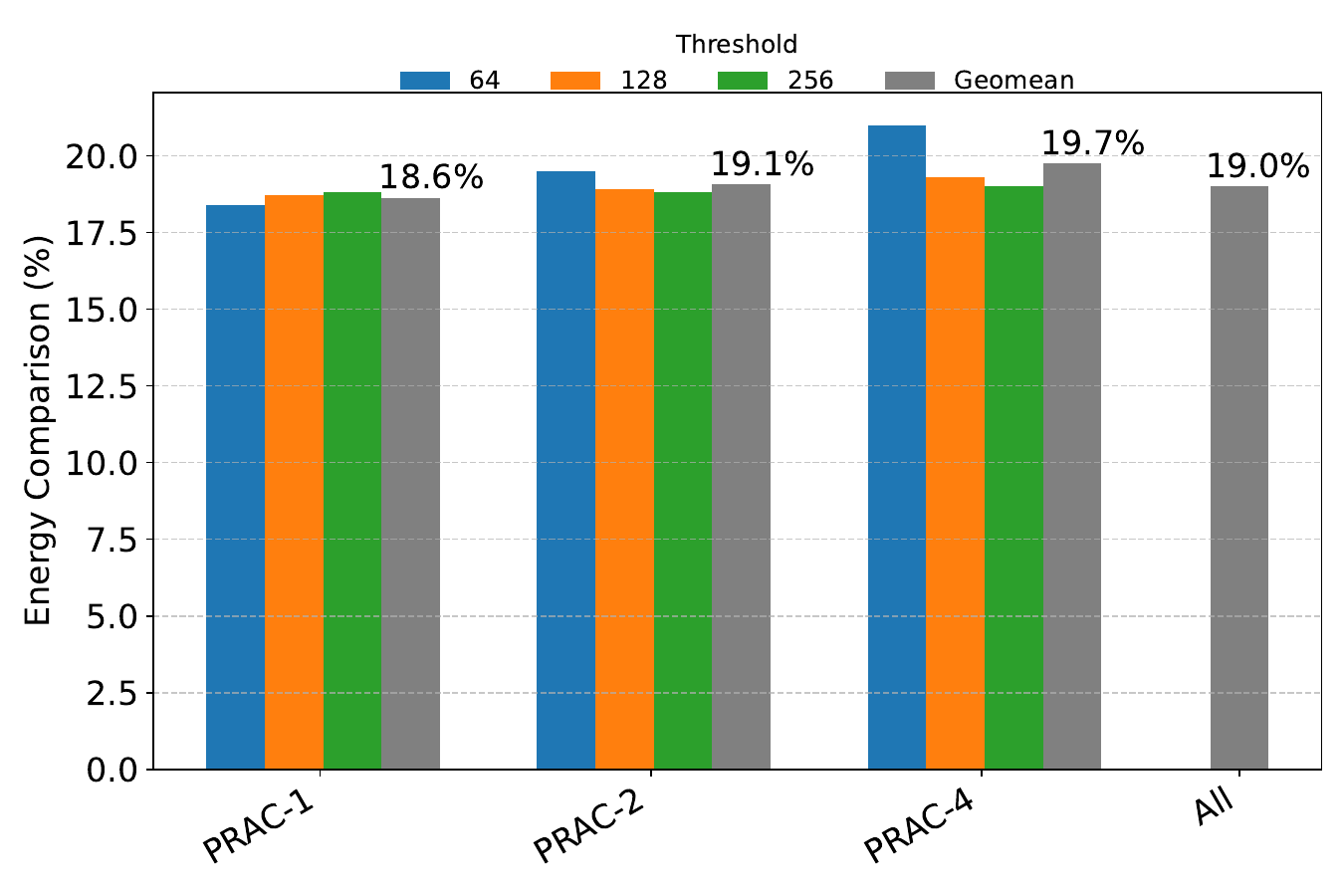}
    \vspace{-16pt}
    \caption{\allchanges{Energy Comparison of \papertitle{} and PRAC for recoveries of 1, 2, and 4 and thresholds of 64, 128, and 256}}
    \label{fig:energy_eval}
\end{figure}

\allchanges{
\textbf{\papertitle{} vs. PRAC+ABO.} The performance evaluation results comparing \papertitle{} with the standard (opportunistic) PRAC+ABO framework are presented in Figure~\ref{fig:opp_prac_grace} and geometric mean results are shown in Figure~\ref{fig:perf_eval_grace_prac}. The evaluation includes both \papertitle{} and PRAC across configurations with n=1,2, and 
4 \texttt{RFM}s as PRAC-1, PRAC-2, and PRAC-4, respectively, with Alert thresholds of 64, 128, and 256. We use a suite of mixed workloads composed of traces with high, medium, and low memory usage characteristics. The results demonstrate that \papertitle{} outperforms the baseline PRAC+ABO framework. Along the mixes, high and medium memory intensity combinations showcase more speedup (up to 20\%).  
Mean values show that \papertitle{} provides 7-9\% better performance. This performance improvement benefits mostly from subarray-level PRAC updates. The overhead was 6\% on average. Hence, \papertitle{} has 2-3\% performance improvement due to bank-level recovery stalling. Although this performance improvement might seem very low, the critical target of this optimization is to eliminate energy inefficiency.

\textbf{Energy Comparison.} In this evaluation, we compare the energy consumption of \papertitle{} and PRAC+ABO with opportunistic mitigation. Figure~\ref{fig:energy_eval} depicts the difference in the energy consumption due to extra RFM refreshes of opportunistic mitigation. The results align with our expectations; while the opportunistic mode of PRAC+ABO improves performance by preemptively refreshing rows before they become hot, it also introduces a large number of unnecessary refreshes, significantly increasing energy consumption.
On average, across all evaluated recovery and threshold configurations, opportunistic PRAC+ABO consumes 19\% more energy compared to \papertitle{}. Across different configurations of recoveries (1,2, and 4) and thresholds, the values slightly vary. The biggest difference happens with PRAC-4 (rec=4), and the threshold 64, which aligns with the observation of the most unnecessary RFM refreshes in sec~\ref{sec:motivation}. These findings underscore a fundamental trade-off: while opportunistic mitigation strategies can minimize performance degradation by anticipating high-activation rows early, this comes at the cost of substantially increased energy usage, reducing the overall efficiency of the memory system.
}

\begin{figure}
    \centering
    \includegraphics[width=1\linewidth]{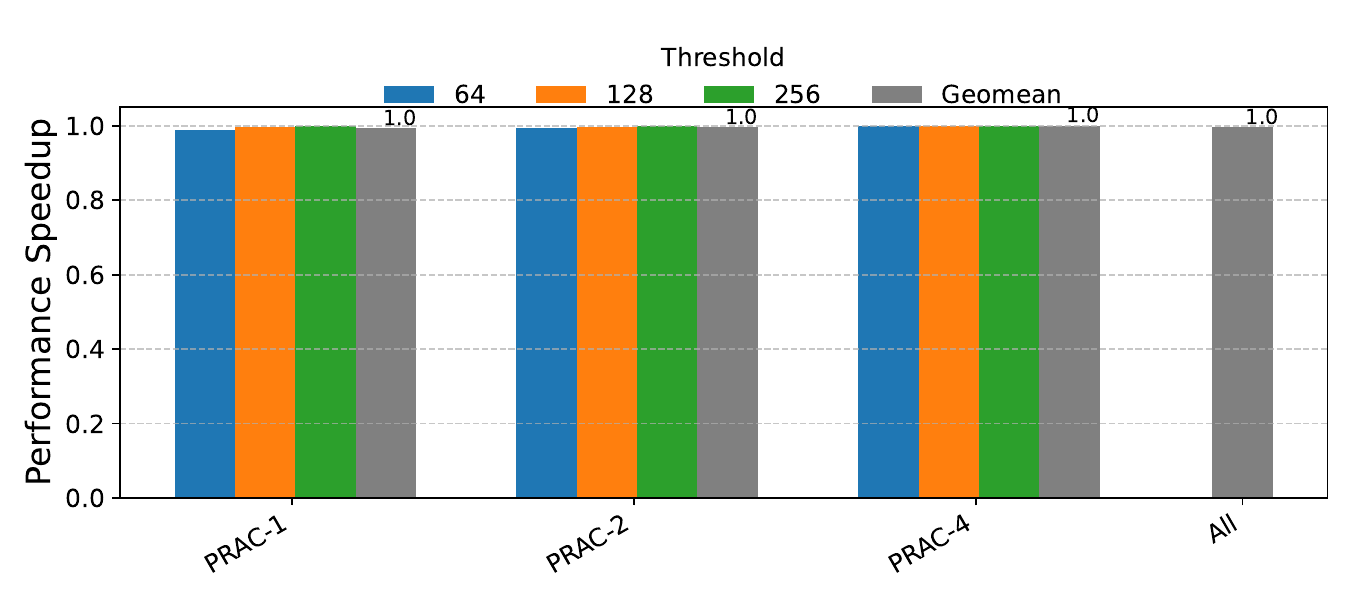}
    \caption{\allchanges{Performance Comparison of \papertitle{} and Baseline architecture for recoveries of 1, 2, and 4 and thresholds of 64, 128, and 256. Baseline architecture refers to no mitigation.}}
    \label{fig:baseline_eval}
    \vspace{-0.5cm}
\end{figure}

\allchanges{
\textbf{\papertitle{} vs Baseline architecture.} 
In this evaluation, we assess the performance of \papertitle{} in comparison to a baseline architecture, where the baseline represents a DRAM system without any Rowhammer mitigation mechanisms. As illustrated in Figure~\ref{fig:baseline_eval}, \papertitle{} achieves performance levels nearly identical to the baseline. The only observable performance degradation is a minor slowdown of approximately 1\%, which occurs under the configuration with a recovery count of 1 or 2 \texttt{RFM}s and a threshold of 64. This minimal overhead demonstrates that \papertitle{} introduces negligible performance penalties, effectively maintaining baseline performance even in the presence of Rowhammer protection.
}

\allchanges{
\textbf{Non-opportunistic  PRAC+ABO vs \papertitle{}.} We have shown that opportunistic mitigation has a wide energy-performance trade-off. In order to save performance, PRAC+ABO can be non-opportunistic, meaning that upon recovery (during \texttt{RFM\textsubscript{ab}}), only the banks, that have rows with activation count equal to greater than Alert threshold, are undergoing recovery procedure. As expected, the banks that have all rows with counters less than Alert threshold will stay idle during this period, downgrading the performance. However, this method ensures only necessary number of refresh operations are performed, saving the energy. The results are shown in Figure~\ref{fig:perf_eval_grace_prac}. Our evaluation considers thresholds of 64, 128, and 256, with a configuration of 1, 2, and 4 \texttt{RFM}s per recovery. Across all combinations of recovery mechanisms (denoted as \texttt{RFM}s) and thresholds, we observe an average performance improvement of approximately 20\%. The lowest performance gain, around 10\%, occurs when the threshold is set to 256 and only one RFM recovery is employed. In contrast, the highest performance improvement is observed under the configuration with a threshold of 64 and one RFM recovery, where performance improves by more than 50\%.
    
}

\begin{figure}
    \centering
    \includegraphics[width=0.9\linewidth]{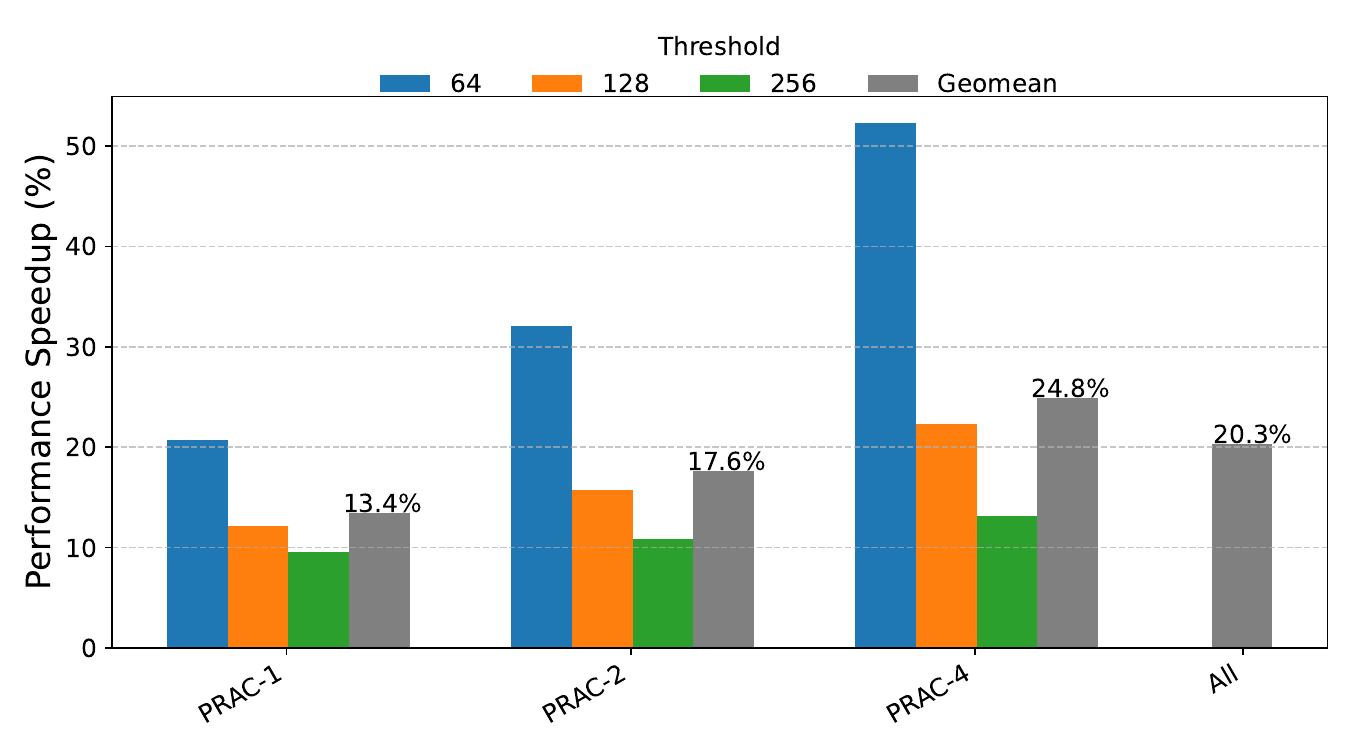}
    \caption{\allchanges{Performance Comparison of \papertitle{} and non-opportunistic PRAC for recoveries of 1, 2, and 4 and thresholds of 64, 128, and 256}}
    \label{fig:nonopp_eval_grace_prac}
\end{figure}

\begin{figure}
    \centering
    \includegraphics[width=1\linewidth]{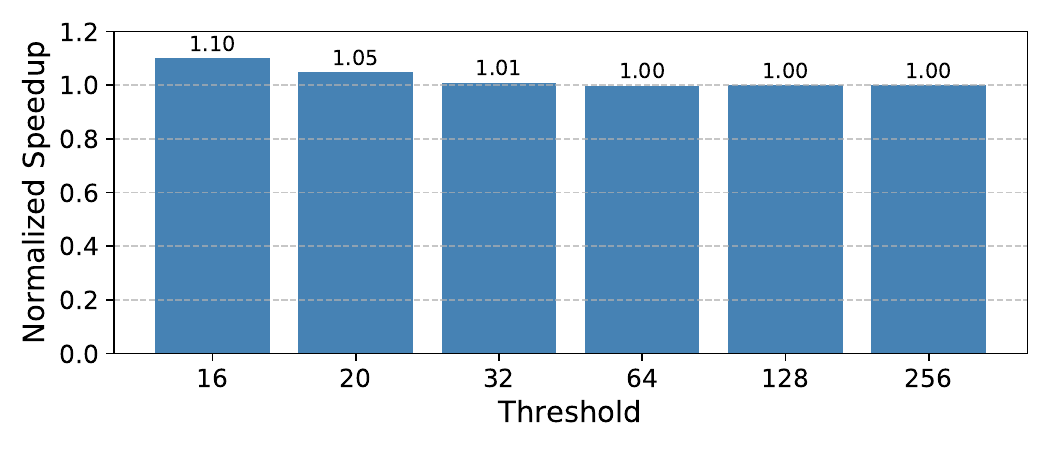}
    \vspace{-0.5cm}
    \caption{\allchanges{Normalized performance of Chronus implemented with \papertitle{} over PRAC+ABO.}}
    \label{fig:chronus_grace_comparison}
    \vspace{-0.5cm}
\end{figure}

\hlgray{
\textbf{Chronus equipped with \papertitle{} vs. PRAC.}
Chronus}~\cite{canpolat2025chronus} \hlgray{optimizes the counter update mechanism in DRAM by relocating per-row activation counters to a dedicated, independent subarray. While this design introduces additional energy overhead, since each row activation results in two separate row accesses (one for the target row and one for the corresponding counter row) and occupies valuable DRAM space, it effectively eliminates the performance penalty typically incurred by in-place counter updates.

In this evaluation, we integrate the \papertitle{} bank stalling mechanism and subarray-level PRAC update into the Chronus architecture and compare the combined approach against the Chronus with an opportunistic PRAC+ABO framework. As shown in Figure}~\ref{fig:chronus_grace_comparison}\hlgray{, the hybrid Chronus+\papertitle{} implementation achieves the same performance as the base implementation with PRAC+ABO.  These results demonstrate that simple and energy efficient \papertitle{} can indeed have the same result as expensive and inefficient base design. Note that even though the \papertitle{} has a small PRAC overhead due to subarray conflicts, this is negligible in complete design. Also note that the Chronus has more energy difference due to double row activation, which incurs an additional  19.07\% energy consumption for each row access. We can conclude that \papertitle{} can achieve same (even better at lower thresholds) performance with very low energy consumption.}

\subsection{Hardware Overhead and Complexity Analysis}
\label{subsec:hardware_overhead}
\allchanges{In this section, we discuss the practicality perspective of the proposed solution, \papertitle{}. Specifically, we discuss the practicality of \papertitle{} in two aspects: changes to JEDEC standards and the area overhead of changes.}

\noindent 
\hlgreen{\textbf{Changes to JEDEC Standards:} 
\papertitle{} requires several architectural modifications to both the DRAM device and the memory controller (MC) interface. The most critical modification is to decouple the increment circuit from the global circuit and connect it to the subarray's local row buffers. Another required modification stems from the fact that MCs typically lack knowledge of the subarray mapping within each DRAM bank. However, this can be addressed with a dedicated register in DRAM that holds subarray mapping information. Then, at boot time, the MC can read these registers and store the corresponding mapping functions in its internal registers for use during execution. Additionally, another hardware extension involves introducing a single register that maintains one bit per bank to track whether the bank needs mitigation. For a 64-bank channel, this mechanism requires only a 64-bit register, making it a low-overhead enhancement.}

\noindent \hlyellow{
\textbf{Area Overhead of Increment Circuit:} \papertitle{} introduces two critical modifications to the DRAM architecture: (1) an increment unit and its control path integrated into local row buffers, and (2) an 
n-bit register to represent the bank mask in an
n-bank DIMM. In our evaluation setup, the DIMM consists of 64 banks, which requires an 8-byte (64-bit) register to track bank-level status.
The first modification decouples the increment logic from the global row buffer and shares it across local row buffers. Since the counter update overlaps with the activation of the next row, it must be routed to the correct subarray. To enable this, we need to have an additional global row address decoder that maps addresses to subarray IDs and row addresses within subarrays, along with an additional comparator in each subarray to resolve subarray ID matches. This ensures that counter updates can be independently delivered, even when the comparator is concurrently needed for the next precharge operation. We evaluate the area overhead of this modification using CACTI}~\cite{balasubramonian2017cacti}\hlyellow{ and Synopsys Design Compiler}~\cite{synopsys_dc}\hlyellow{, and find it to be only 0.03\%, making it negligible.
}

\subsection{Security Evaluation}
\label{sec:sec_eval}


\textbf{Resilience against Memory Slowdown Attacks.}
In Section~\ref{sec3.3}, we stated that the current PRAC+ABO design is vulnerable to MOAT's Torrent-of-Staggered-Alerts attack~\cite{qureshi2024moat}, which exploit the channel-wide stalling behavior of the ABO protocol. Besides, an attacker can trigger an Alert on a single bank to stall the entire memory channel \allchanges{frequently to cause performance attacks}. \papertitle{} mitigates this vulnerability by restricting the Alert-induced stall to only the affected bank. As a result, the effectiveness of \allchanges{performance} attack is significantly diminished, as it relies on a single bank being capable of blocking the entire channel.
\begin{figure}
    \centering
    \includegraphics[width=1\linewidth]{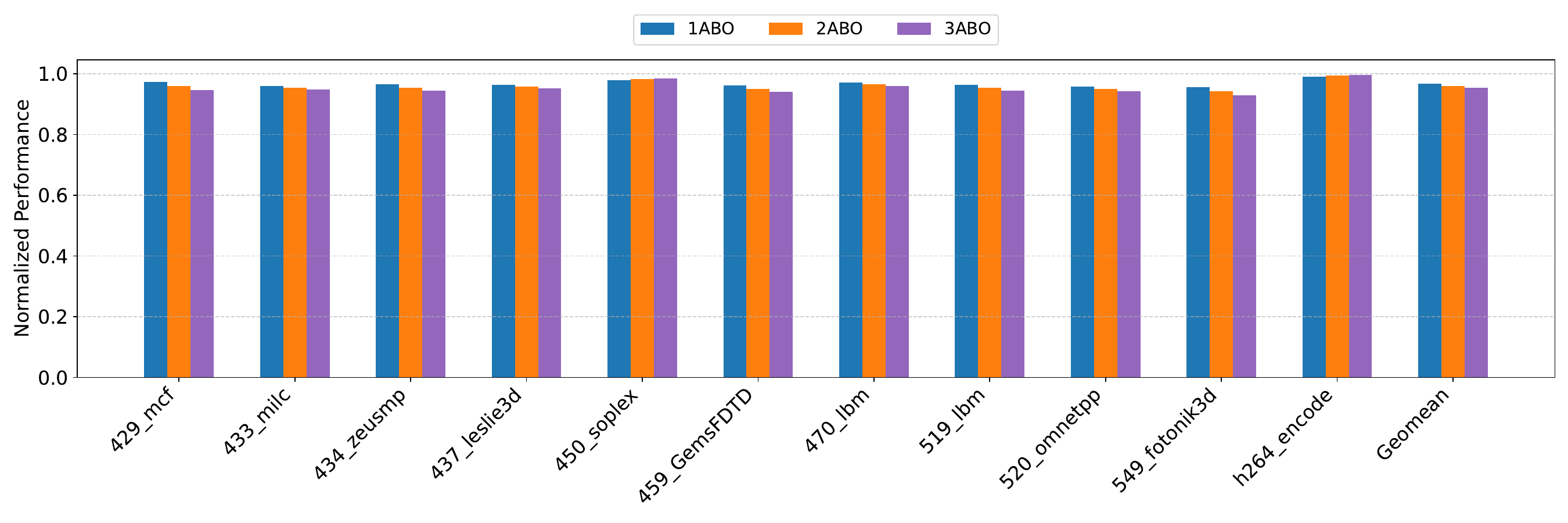}
    
    \caption{\allchanges{Normalized performance of \papertitle{} for each single benign benchmark when there is a performance attack.}}
    
    \label{fig:grace_resilience}
\end{figure}

To evaluate \papertitle{}'s robustness, we repeated the performance attack experiments on a \papertitle{}-enabled system. The results, shown in Figure~\ref{fig:grace_resilience}, use the \textit{motivation} set—consisting of high memory-intensity, single-trace workloads. Under PRAC+ABO, the attack-induced slowdowns reached over 80\%, severely degrading the performance of benign workloads. In contrast, \papertitle{} demonstrates strong resilience to such memory-based performance attacks. The system experiences an average (geometric mean) slowdown of less than 6\%. This is primarily due to the attacker operating on the same banks accessed by these benchmarks. \allchanges{ The most slowdown happens when attacker can induce 3 ABO stalls per \texttt{tREFI}s.  Even in these cases, the maximum observed slowdown remains below 6\%, which is considered a tolerable level for secure memory systems.}

\hlbrown{\textbf{Security of \papertitle{}:} While PRAC+ABO alone does not constitute a complete security solution on its own.  Any secure Rowhammer mitigation framework built upon PRAC+ABO must use a policy to ensure that any DRAM bank exceeding the activation threshold is properly mitigated. From this perspective, \papertitle{} is designed to preserve this baseline security requirement while delivering improved performance and energy efficiency. \papertitle{} targets optimizations in two key components of the PRAC+ABO framework: the per-row activation counter (PRAC) and the Alert-back-off signaling mechanism.   
\papertitle{} counter updates are committed in parallel with subarray accesses by leveraging } \hlbrown{a centralized} \hlbrown{ increment circuitry, in conjunction with a memory controller that maintains subarray mappings. \papertitle{} provides memory with an alert signal to notify MC to stall requests at the bank level. Since we change channel-level blocking to bank-level blocking, we discuss the following potential security issue.

 Since \papertitle{} allows access to non-mitigated banks, this potentially provides an attacker with an opportunity during the recovery for a period to send more activations to a target bank that is close to the Alert threshold. As an example, assume bank A sets its bit in ~\texttt{RFM\_MASK} and sends alert and bank B is in normal operation of serving requests with its highest activation count close to the threshold, }\(T - 1\)\hlbrown{. While the mitigation of bank A is performed, an attacker can send up to }\( N_{\texttt{ACT}} = \frac{t_{\text{\texttt{RFM\_MASK}}}}{t_{\text{RC}}} \)\hlbrown{ \texttt{ACT}s to bank B. }\(N_{ACT}\) \hlbrown{is at most 6} \hlbrown{since the RFM duration is 350ns and \texttt{tRC} is 52ns}\hlbrown{. Therefore, bank B will have rows with at most} \(T + 5\) \hlbrown{activation counts. \papertitle{} decreases the Alert threshold by this maximum value (5 in our system) to account for the worst-case scenario.} 
 \hlbrown{If during recovery, the counter of a row in a bank reaches threshold, it sets the bit and sends the Alert signal. Once this recovery period is finished, in the next recovery, this bank also performs necessary mitigation. Since \papertitle{} uses a safe threshold, it guarantees the same security as the PRAC+ABO guarantees.
 In general, \papertitle{} does not need to lower the threshold. It can keep the ALERT threshold same, and have a second threshold that sets bank bits to 1 in BA register. Once there is row counter reaching the second threshold, its corresponding bank bit is set to 1. Once a row counter reaches the Alert threshold, an alert is sent to MC. For simplicity, we lower the threshold to lower and safe level.}

\section{Related Work}
This paper first eagerly analyzes the efficiency of the existing JEDEC PRAC+ABO using a diverse set of benchmarks and secondly proposes a new solution called \papertitle{} to improve the performance of memory operations while keeping the same security guarantees provided by PRAC+ABO standards. \papertitle{} is a transparent hardware solution and can be deeply related to other works, such as counter-tracking mechanisms, hardware and software mitigations, and so on.

\textbf{Counter-tracking mechanism}. \hlgray{The concept of using activation counters was introduced and patented around by many works}~\cite{bains2016distributed, kim2014architectural,seyedzadeh2016counter,seyedzadeh2018mitigating, lee2019twice, kang2020cat,qureshi2022hydra,woo2023scalable}. Later works ~\cite{park2020graphene, marazzi2022protrr, kim2022mithril, jaleel2024probabilistic} proposed secure in-DRAM tracking mechanisms to tackle energy and performance issues. Graphene~\cite{park2020graphene} designs lightweight RowHammer protection to identify the frequent inputs of incoming stream achieving near-zero performance-energy overhead. 
Mithril ~\cite{kim2022mithril} is the first to propose DRAM-memory controller cooperative mitigation using in-DRAM tracking. Aamer Jaleel et al. ~\cite{jaleel2024probabilistic} proposes to manage trackers with probabilistic management
policies like request-stream sampling and random evictions. PROTRR \cite{marazzi2022protrr} uses frequent item counting to track aggressor rows in DRAM. 


\textbf{DRAM Performance Mechanisms.} \hlgray{Subarray-level parallelism has been investigated in several prior studies. Kim et al.}~\cite{kim2012case} \hlgray{proposed subarray-level parallel memory access mechanisms to improve memory bandwidth and performance. Hassan et al.}~\cite{hassan2024self} \hlgray{introduced a self-managing DRAM architecture that leverages the independence of subarrays to enable autonomous management within DRAM devices. Our approach is to eliminate the obscurity for subarray parallelism due to the increment circuit design in PRAC. Specifically, we propose modification of the circuit to enable overlapping only the counter update phase rather than the entire precharge phase with the subsequent activation, thereby minimizing performance overhead. Additionally, Chang et al.~}\cite{6835946} \hlgray{explored subarray-level refresh operations, while HiRa}~\cite{hira} \hlgray{presents methods to reduce refresh latency by concurrently refreshing two rows connected to distinct charge restoration circuitries.} BreakHammer ~\cite{canpolat2024breakhammer} removes the performance overhead of Rowhammer attacks by identifying and throttling hardware threats that frequently triggter preventive actions.

\textbf{Slowdown Attacks.}  Due to the impact of DRAM on performance, the research community has long explored threats that exploit shared memory subsystems to degrade performance, often in the form of Denial-of-Service (DoS) attacks~\cite{mutlu2007memory}. While recent Rowhammer mitigation mechanisms aim to enhance system reliability, they can inadvertently introduce substantial performance overheads, particularly when relying on aggressive repair strategies such as frequent refreshes ~\cite{bhati2015flexible, mukundan2013understanding} or row remapping ~\cite{saxena2024rubix}.  \allchanges{Different works~\cite{nazaraliyevnot,woo2025dapper,canpolat2024understanding} describe ways to exploit RH mitigation for performance and side channel attacks.}
\section{Conclusion}

In this paper, we tackle the performance and energy overhead inherent in PRAC+ABO. We propose \papertitle{}, a novel PRAC+ABO enhancement featuring a two-level optimization that enables subarray-level counter updates and allows DRAM banks to mitigate RowHammer independently without stalling. \papertitle{} introduces minimal hardware changes—namely, centralized increment circuit connected to subarray and a single global register, called bank-alert register, where each bit indicates whether a specific bank needs a mitigation — enabling the memory controller to continue serving requests to unaffected banks. Our evaluations show that \papertitle{} improves performance by \allchanges{a geomean of 8\% and saves energy by an average of 20\%  over PRAC+ABO. Its performance is same as the baseline - no mitigation architecture and energy consumption is minimal}. Overall, \papertitle{} provides an efficient and practical enhancement to PRAC+ABO, balancing performance and security with low hardware overhead.

\nocite{*} 
\bibliographystyle{ACM-Reference-Format}
\bibliography{ref}

\end{document}